\documentclass[12pt]{article} 

\usepackage{bbding}
\usepackage{mathtools}
\PassOptionsToPackage{normalem}{ulem}
\usepackage{ulem}

\usepackage{tikz,tikz-3dplot}
\usepackage{xcolor}
\usetikzlibrary{positioning}
\usetikzlibrary{3d}

\usepackage{epsfig}
\usepackage{graphicx}
\usepackage{comment}
\usepackage{latexsym}
\usepackage{hyperref}
\usepackage{amsmath}
\usepackage{color}
\usepackage{amsbsy}
\usepackage{amssymb}
\usepackage{amsthm}
\usepackage{amsfonts}
\usepackage{cite}
\usepackage{enumitem}

\newcommand{\F}{{\cal F}}
\newcommand{\N}{{\cal N}}
\newcommand{\V}{{\cal V}}
\newcommand{\G}{{\cal G}}
\newcommand{\W}{{\cal W}}
\newcommand{\T}{{\cal T}}
\newcommand{\K}{{\cal K}}
\newcommand{\A}{{\cal A}}
\newcommand{\M}{{\cal M}}
\newcommand{\Hc}{{\cal H}}
\newcommand{\nF}{n_{F}^{(0)}}
\newcommand{\nB}{n_{B}^{(0)}}
\newcommand{\half}{{1\over 2}}
\renewcommand{\SS}{Scherk-Schwarz }

\newcommand{\be}[0]{\begin{equation}}
\newcommand{\ee}[0]{\end{equation}}
\newcommand{\dis}{\displaystyle}

\newcommand{\Z}{\mathbb{Z}}
\newcommand{\integer}{\mathbb{N}}

\renewcommand{\O}{{\cal O}}

\newcommand{\diag}{{\rm diag}\,}

\newcommand{\Str}{\textrm{Str}\,}

\newcommand{\ie}{{\em i.e.} }
\newcommand{\eg}{{\em e.g.} }

\newcommand{\where}{\mbox{where}}

\renewcommand{\and}{\mbox{and}}

\newcommand{\esps}{\phantom{\!\!\!\overset{|}{a}}}
\newcommand{\esp}{\phantom{\!\!\overset{\displaystyle |}{|}}}
\newcommand{\espD}{\phantom{\!\!\underset{\displaystyle |}{\cdot}}}

\newcommand{\bm}{\boldmath} 

\catcode`\@=11
\def\marginnote#1{}
\newcount\hour
\newcount\minute
\newtoks\amorpm
\hour=\time\divide\hour by60 \minute=\time{\multiply\hour by60
\global\advance\minute by-\hour}
\edef\standardtime{{\ifnum\hour<12 \global\amorpm={am}%
        \else\global\amorpm={pm}\advance\hour by-12 \fi
        \ifnum\hour=0 \hour=12 \fi
        \number\hour:\ifnum\minute<10 0\fi\number\minute\the\amorpm}}
\edef\militarytime{\number\hour:\ifnum\minute<10 0\fi\number\minute}
\def\draftlabel#1{{\@bsphack\if@filesw {\let\thepage\relax
   \xdef\@gtempa{\write\@auxout{\string
      \newlabel{#1}{{\@currentlabel}{\thepage}}}}}\@gtempa
   \if@nobreak \ifvmode\nobreak\fi\fi\fi\@esphack}
        \gdef\@eqnlabel{#1}}
\def\@eqnlabel{}
\def\@vacuum{}
\def\draftmarginnote#1{\marginpar{\raggedright\scriptsize\tt#1}}
\def\draft{\oddsidemargin -.2truein
        \def\@oddfoot{\sl preliminary draft \hfil
        \rm\thepage\hfil\sl\today\quad\militarytime}
        \let\@evenfoot\@oddfoot \overfullrule 3pt
        \let\label=\draftlabel
        \let\marginnote=\draftmarginnote
   \def\@eqnnum{(\theequation)\rlap{\kern\marginparsep\tt\@eqnlabel}%
\global\let\@eqnlabel\@vacuum}  }
\def\thebibliography#1{
\vskip 0.5cm \centerline{\bf \Large References}
\list{
[\arabic{enumi}]}{\settowidth\labelwidth{[#1]}
\leftmargin\labelwidth
\advance\leftmargin\labelsep
\usecounter{enumi}}
\def\newblock{\hskip .11em plus .33em minus .07em}
\sloppy\clubpenalty4000\widowpenalty4000
\sfcode`\.=1000\relax}

\renewcommand{\theequation}{\arabic{section}.\arabic{equation}}
\renewcommand{\section}{\setcounter{equation}{0}\@startsection
{section}{1}{0mm}{-\baselineskip}{0.5\baselineskip} {\normalfont\Large\bfseries}}
\renewcommand{\subsection}{\@startsection
{subsection}{2}{0mm}{-\baselineskip}{0.5\baselineskip} {\normalfont\large\bfseries}}
\renewcommand{\subsubsection}{\@startsection
{subsubsection}{3}{0mm}{-\baselineskip}{0.5\baselineskip}
{\normalfont\normalsize\slshape}}

\begin{document}


\begin{titlepage}
\begin{flushright}
CPHT-RR117.122018, DCPT-18/37, IPPP/18/112, December 2018
\end{flushright}

\vspace{1cm}

\begin{centering}
{\bm\bf \Large Stability and vacuum energy in }\\
\vspace{2mm}
{\bm\bf \Large  open string models with broken supersymmetry   }\\

\vspace{12mm}
 {\bf Steven Abel$^{1,2}$, Emilian Dudas$^{3}$\\
 Daniel Lewis$^2$ and Herv\'e Partouche$^3$}

\vspace{2mm}

$^1$ Institute for Particle Physics Phenomenology, Durham University, and \\ 
$^2$ Department of Mathematical Sciences,\\ South Road, Durham, U.K.\\
{\em  s.a.abel@durham.ac.uk, daniel.lewis@durham.ac.uk}

\vskip .1cm

$^3$ {Centre de Physique Th\'eorique, Ecole Polytechnique, CNRS$^\dagger$,
\\
F-91128 Palaiseau, France\\
{\em emilian.dudas@polytechnique.edu, herve.partouche@polytechnique.edu}} 
 \vspace{10mm}

{\bf\Large Abstract}

\end{centering}


\noindent 

{\noindent 

We construct type~I string models with supersymmetry broken by compactification that are non-tachyonic and  have exponentially small effective potential at one-loop.    
All open string moduli can be stabilized, while the closed string moduli remain massless at one-loop. 
The backgrounds of interest have rigid Wilson lines by the use of stacked branes, and some models should have heterotic duals.
We also present non-tachyonic backgrounds with positive potentials of runaway type at one-loop. This class of models could be used to test various swampland conjectures.} 

\vskip 1cm

\vspace{5pt} \vfill \hrule width 6.7cm \vskip.1mm{\small \small \small \noindent 
$^\dagger$\ Unit{\'e} mixte du CNRS et de l'Ecole Polytechnique,
UMR 7644.}

\end{titlepage}
\newpage
\setcounter{footnote}{0}
\renewcommand{\thefootnote}{\arabic{footnote}}
 \setlength{\baselineskip}{.7cm} \setlength{\parskip}{.2cm}

\setcounter{section}{0}


\section{Introduction}

This paper explores new (geometric) methods of constructing string theories with spontaneously broken supersymmetry that have enhanced stability, and conceivably naturalness, as a possible route to a string embedding of the Standard Model. 

In the past decades, in order to discover important properties and ingredients of string theory such as dualities and branes, exactly supersymmetric models have been the focus of attention, and have been considered from various points of view. Compactifications on Calabi-Yau manifolds~\cite{SStringsIntr}, 
orbifold models~\cite{orbifolds},  
fermionic constructions~\cite{Kawai:1986ah,Antoniadis:1986rn,Kawai:1987ew}, 
or Gepner points in moduli space~\cite{gepner,gepner2} have all been analyzed extensively. 
A key point in all these studies is that supersymmetry
guarantees stability and all-orders consistency in a Minkowski background,
synonymous with the fact that the cosmological constant is precisely
zero in the vacuum. 

Supersymmetry must however be broken to make contact with particle phenomenology and cosmology, and it is natural to consider performing this already at the string level, rather than postponing it to the supersymmetric effective field theory. 
In a theory that has supersymmetry broken at the string scale $M_s$ without tree-level tachyons \cite{so(16)xso(16),Brane_susy_break,reviews}, the quantum effective potential in $D$ dimensions is naturally of order $M_s^D$ \cite{dienes}. This may imply the existence of scalar field tadpoles responsible for the destabilisation of the initial background, with either runaway behaviour or perhaps attraction to AdS-like  vacua associated with very large negative cosmological constants \cite{Angelantonj:2006ut} (it is  not clear if the latter are supersymmetric or not). Consequently there has been continued interest in non-supersymmetric theories where 
firstly the effective potential happens  to cancel at leading or even higher order \cite{L=0,Shiu:1998he,HarveyL=0,1-L=0,Angelantonj:1999gm,Angelantonj:2004cm}, and secondly where 
supersymmetry breaking is under parametric control because the theory lies on an interpolation from an entirely supersymmetric theory \cite{Itoyama:1986ei,open1_ss,FR}.
Recently there has been a resurgence of interest in the fact that there is a large class of theories of the latter kind that have exponentially suppressed 1-loop effective potential \cite{Abel:2015oxa,Satoh:2015nlc,Iengo:1999sm,SNSM,SNSM2,Sugawara:2016lpa,Abel:2017vos,CFP,CP,Partouche:2018ftj,GrootNibbelink:2017luf,Abel:2017rch}.

In the present work, we consider these questions in open string models, where geometric reasoning makes the physical picture much clearer. In particular this will allow us to focus on an issue that has been somewhat neglected in the literature, namely that, even in theories that have vanishing or exponentially small effective potentials, some of the moduli  will generically acquire tachyonic masses at 1-loop~\cite{SNSM,SNSM2,CP}.  By developing a global geometric picture of the potential, one can decide which theories are not unstable: as we shall see, systems that have no such tachyons at 1-loop exist, but are extremely constrained. 

We will study string models where  supersymmetry (with 16 supercharges, implying all moduli to be Wilson lines) is spontaneously broken by coordinate-dependent compactification, which is nothing other than the Scherk-Schwarz mechanism \cite{SS} of field theory, upgraded to string theory: it was developed for closed strings in \cite{SSstring, Kounnas-Rostand} and for open strings in  \cite{open1_ss,open2_ss,open3_ss}. In this context, the scale $M$ of supersymmetry breaking is of order $M_s/R$, where $R$ is the characteristic radius\footnote{Throughout this paper, the radii and moduli are dimensionless. All dimensionful quantities are dressed with appropriate powers
of $M_s$.} of the internal space involved in the mechanism. When $R$ is moderately large, the leading contribution to the 1-loop effective potential $\V$  is Casimir-like, and arises from the Bose-Fermi non-degeneracy of the shift in the Kaluza-Klein (KK) towers. As a rule-of-thumb, assuming that  the lightest mass scale of the background is $M$, the effective potential is, up to exponentially suppressed terms\footnote{These terms are $\O(e^{-cM_s/M})$, where $c=\O(1)$. When $M$ is at least 3 orders of magnitude lower than $M_s$, they are (much) lower than $~10^{-120}M_s^D$, which is the order of magnitude of the observed cosmological constant in dimension 4. Therefore, they  have no observational consequences and can be safely omitted.}, 
\be
\label{Vn}
\V~\simeq~\big(\nF-\nB\big) \, \xi_D \, M^D~.
\ee
This dominant contribution arises from the $\nF$ and $\nB$ massless fermionic and bosonic degrees of freedom of the model, together with their light KK towers of modes accounted by the overall dressing $\xi_D>0$. As explained below, such backgrounds yield critical points of the potential,  with respect to all Wilson line (WL) deformations. Thus,  the rule-of-thumb is not a replacement for the full potential -- it corresponds to just the first term in a Taylor expansion in WLs about the critical point -- but it can be used to compare the potential energy at different critical points. Criticality relies on the next term vanishing i.e., denoting the gauge group by $\G$ and the WLs by $a_r^I$,\footnote{\label{foo}Contrary to Eq.~(\ref{Vn}), there is an equal sign in Eq.~(\ref{dV0}) that   follows from an {\em exact symmetry}, which is a gauge symmetry. This exact vanishing is expected to be valid to all order in perturbation theory \cite {GV}.}   
\be
\label{dV0}
{\partial \V\over \partial a^I_r}~=~0\, , \quad { I\,=\,D},\dots,9\, , \quad r\,=\,1,\dots, \text{rank}\,\G~.
\ee
In Ref.~\cite{GV}, this is achieved at points of enhanced symmetry, where states with non-trivial Cartan charges are massless.\footnote{Denoting $Q_r$ the charge operator associated with the $r$-th Cartan $U(1)$, $\V$ contains at linear order in $a_r^I$ a contribution $\sum Q_r a_r^I$, where the sum is over the massless spectrum. Vanishing of this tadpole follows from the fact that any state of charge $Q$ can be paired with a state of charge $-Q$.} In our context, such a  massless state belongs to a KK tower of modes with masses $2kM$, $k\in\Z$, and identical spin, while the superpartners have masses $(2k+1)M$. When an extra massless state acquires a mass by switching on WL's, we leave criticality till the mass reaches $M$. The reason for this is that one mode in the KK tower of superpartners is now massless. The latter having non-trivial Cartan charges, we regain criticality.\footnote{Tadpoles can be analyzed by switching on WLs one by one. See e.g. Eq.~(\ref{shift}), with $a_\beta=0$ and let $a_\alpha$ varying from 0 to $\half$. An extra  massless boson $(m_9,F)=(0,0)$ at $a_\alpha=0$ is replaced by an extra massless fermion $(m_9,F)=(-1,1)$ at $a_\alpha=\half$. When $0<a_\alpha<\half$, we leave criticality and a mass scale exists in the range $(0,M)$.} In other words, Eq.~(\ref{dV0})  is valid provided there is no massive particle with mass less than $M$. 
Under this assumption, when the background satisfies non-supersymmetric Bose-Fermi degeneracy at the massless level, $\nF=\nB$, not only do the dominant contribution to the 1-loop potential vanishes, but also no tadpole survives at all, including those of the supersymmetry breaking scale $M$ and the dilaton fields. (In this case, at the 1-loop level,  the only possible tadpole which is for  $M$ is exponentially suppressed.)

Eq.~\eqref{Vn} allows some simple general statements to be made {\it a priori}. For example, in the absence of any open string WL deformation, supersymmetry breaking \`a la Scherk-Schwarz is known to give an excess of massless bosons, so the scalar potential is negative. However WLs then provide a simple
method for increasing the potential. We can exemplify this in the simplest realisation, which is the nine-dimensional case. Let us add a Wilson line on a D9-brane that corresponds, after a T-duality on the circle of radius $R_9$, to 
a D8-brane sitting in the other fixed point $\pi \tilde R_9\equiv \pi/R_9$ of the orientifold operation $\Omega' = \Omega \Pi$, where $\Pi$ is the parity on the dual circle coordinate. In the absence of supersymmetry breaking, open strings 
stretched between the D8-branes at the origin and the one at $\pi \tilde R_9$ have masses, before T-duality, given by $(m_9+\half)M_s/R_9$. Supersymmetry breaking adds an additional shift of $\half$ in the fermion masses,
such that the fermions stretched between the stack of branes at the origin and the one at $\pi \tilde R_9$ become massless. Compared to the case where all D8-branes are at the origin with maximal gauge group $SO(32)$, this new configuration,
with a gauge group that we denote  $SO(31) \times SO(1)$ has a  higher scalar potential. The configuration is also stable, as will be shown in the next sections. 
Note the well-known fact that this Wilson line is in $O(32)$ but not in $SO(32)$.  It is also not a continuous Wilson line, but rather a discrete one, which signifies that the brane at $\pi \tilde R_9$ is not a regular brane, in the sense that it has no position moduli. It is rather a half-brane stuck at the fixed point, with no associated gauge group. 
In nine dimensions, beside the $SO(32)$ case, this is the only stable configuration and both of them have negative scalar potential. However, by suitable further compactification and by distributing other D-branes at different orientifold fixed points, one can construct stable configurations with zero or positive effective potentials, with Wilson lines in either $SO(32)$ or $O(32)$.

Of course, to achieve vanishing of the effective potential (at 1-loop, and up to exponentially suppressed terms) and stability of the background, the mass-squared terms of the WLs must be non-negative. However, finding backgrounds satisfying $\V\ge 0$ without tachyonic moduli proves to be rather delicate. This is due to the fact that the WL masses $m_r^I$ depend on the difference between the Dynkin indices  $T_{{\cal R}_B}$ and $T_{{\cal R}_F}$ of the representations of the massless bosons and fermions~\cite{SNSM,SNSM2,CP}~:
\be
(m_r^I)^2 ~\simeq~ (T_{{\cal R}_B}-T_{{\cal R}_F})\xi^{\prime\prime},\quad \where \quad \xi^{\prime\prime}~>~0~,~ \quad T_{\cal R}\delta_{ab}~=~{1\over 2}\text{tr}\, T_aT_b~,
\ee
and  the $T_a$'s are Hermitian generators in the representation ${\cal R}$. As a result, one can see that generally the more positive $\V$ is, the more unstable the background is, because the massless fermions that contribute positively to the potential energy, also contribute negatively to the WL mass-squared. Therefore it is non-trivial that such stable backgrounds exist.  Furthermore, the notion of stability of the universe itself can be addressed from a cosmological point of view. As shown in Refs~\cite{CFP,CP,Partouche:2018ftj}, it turns out that flat, expanding universes are way more natural when $\nF-\nB\ge 0$, due to an attractor mechanism towards a so-called  ``Quantum No-Scale Regime'', which is characterized by evolutions converging to those found for $\nF-\nB= 0$.

This paper develops a systematic and geometric approach to constructing  
{\em backgrounds with effective potential vanishing  at critical points and non-tachyonic,  at the 1-loop level and up to exponentially suppressed terms.} This is done within open string theories, which in principle
contain the duals of heterotic theories. We also find backgrounds with $\nF-\nB>0$ that are non-tachyonic at 1-loop and where $M$ slides to low supersymmetry breaking scale along its positive potential. One may ask why constructing such models with positive sign of the scalar potential, or with leading term absent, and tachyon free at 1-loop may be relevant. The question is valid since, even when $\nF-\nB=0$,  some of the moduli, and in particular the supersymmetry breaking scale $M$,  are not stabilized and one cannot talk about cosmological constant. 
Our motivations for constructing such models can be summarized in short as follows: 

$\bullet$  Models with positive scalar potential  could be a starting point for constructing quin\-tessence models in string theory. The examples provided in the present work are probably not realistic since the leading potential term is too steep and does not lead to an accelerating universe.  It is however reasonable to believe 
that in more refined models with several contributions to the scalar potential,  some regions in fields space are flat enough and could lead to quintessence.

$\bullet$  Models with vanishing leading term in the 1-loop scalar potential could be a starting point for deriving a suppressed cosmological constant. By adding several perturbative contributions
to the scalar potential, it is in principle  possible to stabilize $M=\O(M_s/R)$  at a small value. At such an extremum, these contributions of the potential will have similar values, therefore of the order of the 1-loop contribution, which is exponentially suppressed for large enough radius. 
 It would be clearly of great interest to construct explicit phenomenological models along these lines. This is however beyond the scope of the present paper. We would like to re-emphasize that all previous constructions of this type in the literature have tachyonic instabilities at 1-loop. Eliminating these instabilities is the main result of our paper.
The fate of such models beyond 1-loop is currently unknown.

In Sect.~\ref{warmup}, we discuss the simplest setup in nine dimensions and the emergence of massless fermions for specific values of the Wilson lines. We also show that  the only stable configurations are $SO(32)$ and $SO(31) \times SO(1)$. 

In Sect.~\ref{stabi}, we discuss compactifications to lower dimensions. By distributing frozen half-branes on more orientifold fixed points, there are more possibilities to obtain massless fermions, because of the interplay of the Scherk-Schwarz supersymmetry breaking with the Wilson lines: the net effect is an increase in the effective potential. Our main result is that a variety of stable brane configurations exist in various dimensions, for which the models are non-tachyonic at 1-loop, even when the scalar potential is positive or vanishing. This statement, which is valid up to exponentially suppressed terms, includes all moduli: while open string WLs acquire masses at the quantum level, the NS-NS and RR moduli arising from the closed string sector are massless at 1-loop (except $M$ which is running away when $\nF-\nB\neq0$). In particular, we find 1-loop marginally stable backgrounds with exponentially small effective potential in 4 dimensions. Notice that the NS-NS moduli (except $M$ and the dilaton) and the RR moduli are expected to be stabilized at 1-loop~\cite{SNSM, SNSM2,CP,Bourliot:2009ai, Estes:2010sh} in the dual heterotic models~\cite{APP}. This is due to the fact that in heterotic string, the internal metric and antisymmetric tensor (dual to the NS-NS metric and RR moduli) are treated on equal footing with the $SO(32)$ Wilson lines present in both theories. In type I, the stabilization of the NS-NS metric and RR 2-form moduli should arise at points in moduli space where nonperturbative D-strings (dual to the heterotic F-string) wrapped in the internal space yield extra massless bosons (see Ref.~\cite{Estes:2011iw} for an analogous effect due to finite temperature). Hence, type~I models with $\nF-\nB> 0$ are required for their dual heterotic descriptions to have all moduli except $M$ and the dilaton stabilized at 1-loop, with $\nF-\nB$ evaluated on the heterotic side vanishing~\cite{APP}. 

In Sect.~\ref{sec.4}, we discuss nonperturbative aspects of these models,  by identifying whether the Wilson lines belong to $SO(32)$ or $O(32)$.  Since the latter do not have heterotic duals, they do not
exist nonpertubatively.  

Sect.~\ref{sec.5} contains comments about the relevance of our constructions to, and their compatibility with, the various swampland conjectures \cite{Vafa:2005ui,ArkaniHamed:2006dz, Ooguri:2006in,Ooguri:2016pdq,Obied:2018sgi,Agrawal:2018own,Garg:2018reu,Ooguri:2018wrx}.

Finally, our conclusions and perspectives can be found in Sec.~\ref{sec.6}. They are followed by rather extensive appendices, collecting the conventions and notations  and the main techniques used to calculate the scalar potentials used throughout the paper. 


\section{No WL stability with \bm $\nF\ge \nB$ in 9 dimensions}
\label{warmup}

In order to find non-tachyonic configurations in non-supersymmetric open
string theory at 1-loop, we will analyse toroidal compactifications of type
I  down to $D$ dimensions and implement a Scherk-Schwarz mechanism. As a warm up, the present section focuses on the simple case of $D=9$, with supersymmetry  broken spontaneously along $S^1(R_9)$, the internal circle of radius $R_9$. 


\subsection{General setup}
\label{GS}

In order to avoid a Hagedorn-like tachyonic instability, we assume $R_9$ to be larger than the Hagedorn radius  $R_H\,=\, \sqrt{2} $\,. Restricting further to values moderately larger than $R_H$ greatly simplifies  the expression for the effective potential, which takes a universal form dominated by the contributions of the pure KK modes. In the closed string sector, at zero winding number $n_9$ along the compact direction $X^9$, as well as in the open string sector, the stringy Scherk-Schwarz mechanism induces a shift of the KK masses according to the fermionic number $F$, which defines the scale $M$ of supersymmetry breaking, 
\be
 {m_9\over R_9}\,M_s~\longrightarrow~  {m_9+{F\over 2}\over R_9}\,M_s\quad~\Longrightarrow\quad~  M~=~{M_s\over 2R_9}~.
\ee
In the open string sector, WL deformations along $S^1(R_9)$ can be introduced, which spontaneously break the gauge group. Considering first the case of the $SO(32)$ theory, the WL matrix living in the Cartan subgroup can be parametrized \be
\begin{aligned}
\W & ~=~ \diag\big(e^{2i\pi a_\alpha}, \alpha=1,\dots, 32\big) \\
&~\equiv~ \diag\big(e^{2i\pi a_1},e^{-2i\pi a_1}, e^{2i\pi a_2},e^{-2i\pi a_2},\dots, e^{2i\pi a_{16}},e^{-2i\pi a_{16}}\big)~ ,  
\end{aligned}
\label{Weven}
\ee
and the open strings having Chan-Paton charges at both ends have their KK masses shifted further  as 
\be
 {m_9\over R_9}\,M_s~\longrightarrow~  {m_9+{F\over 2}+a_\alpha-a_\beta\over R_9}\,M_s~.
 \label{shift}
\ee
It is convenient to T-dualize $S^1(R_9)$ to switch to a geometrical setting  in type~I', with D8-branes and two O8-planes located at $\tilde X^{9}=0$ and $\tilde X^{9}=\pi \tilde R_{9}$, where $\tilde R_9=1/R_9$. In this picture, the 32 $\half$-branes are located at $2\pi a_\alpha \tilde R_9$. The allowed configurations consist of $p_1\in2\integer$ $\half$-branes sitting on the O8-plane at $a=0$, $p_2\in2\integer$ $\half$-branes coincident with the second O8-plane at $a=\half$, and stacks of $r_\sigma$ branes each located at some $a\in \, (0,\half)$, together with their mirrors at $-a$. The gauge symmetry is  $U(1)^2_{G,C} \times SO(p_1)\times SO(p_2)\times \prod_\sigma U(r_\sigma)$, where the mutiplicities are constrained by the RR tadpole condition $p_1+p_2+2\sum_\sigma r_\sigma=32$, and $U(1)_G$, $U(1)_C$ are the Abelian factors generated by the dimensionally reduced metric and RR 2-form, $G_{\mu9}$, $C_{\mu9}$.

\begin{figure}
\begin{center}
\tdplotsetmaincoords{70}{80}    
\begin{tikzpicture}[ 
	tdplot_main_coords, >=stealth, cube/.style={very thick,black}, grid/.style={very thin,gray}, axis/.style={->,blue,thick} ] 

\tikzset{facestyle/.style={fill=green,opacity=0.4,draw=black,very thick,line join=round}}
\tikzset{facestyleg/.style={fill=blue,opacity=0.4,draw=black,very thick,line join=round}}

\begin{scope}[canvas is yx plane at z=0] 
	\draw[red,thick]  (0,0) ellipse (5cm and 5cm); 
	\node at (-5,0cm) [circle,fill=black,opacity=0.5,shading=ball](ball) {}; 
	\node at (5,0cm) [circle,fill=black,shading=ball](ball) {}; 
\end{scope}

\foreach \i in {-1,...,1} \tdplotsetrotatedcoords{0}{0}{\i*5} \draw[tdplot_rotated_coords, canvas is zy plane at x = 0, facestyle] (-1cm,-4.2cm) rectangle (1cm,-5.8cm);
\foreach \i in {-1,...,1} \tdplotsetrotatedcoords{0}{0}{90+\i*2} \draw[tdplot_rotated_coords, canvas is zy plane at x = 0, facestyleg] (-1cm,-4.2cm) rectangle (1cm,-5.8cm);
\foreach \i in {-1,...,1} \tdplotsetrotatedcoords{0}{0}{-90+\i*2} \draw[tdplot_rotated_coords, canvas is zy plane at x = 0, facestyleg] (-1cm,-4.2cm) rectangle (1cm,-5.8cm);
\foreach \i in {2,...,-2} \tdplotsetrotatedcoords{0}{0}{180+\i*4} \draw[tdplot_rotated_coords, canvas is zy plane at x = 0, facestyle] (-1cm,-4.2cm) rectangle (1cm,-5.8cm);

\node[text width=3cm] at (-6.5cm,-0.3cm) {$p_1$ $\half$-branes\\ at $a=0$}; 
\node[text width=3cm] at (7.6cm,0.3cm) {$p_2$ $\half$-branes\\ at $a=\frac{1}{2}$}; 
\node[text width=4cm] at (1.6cm,2.4cm) {$q$ branes images \\at $a=-\frac{1}{4}$}; 
\node[text width=4cm] at (3.4cm,-2.5cm) {$q$ branes \\ at $a=\frac{1}{4}$};

\end{tikzpicture} 
\end{center}
\caption{\label{a1/4}\em \footnotesize A D9-brane configuration in the T-dual picture, in which the WLs become the positions of D8-branes along $\tilde X^9$. In the example depicted, the gauge group is $U(1)_{G,C}^2\times SO(p_{1})\times SO(p_{2})\times U(q)$.}
\end{figure}
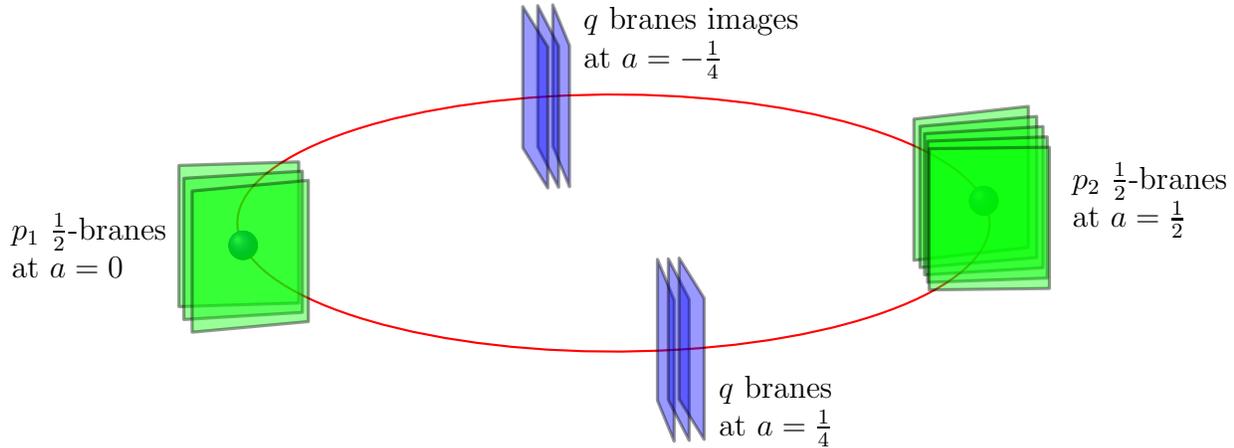

Ultimately, we wish to let the branes move anywhere and find their final stable configurations, possibly with positive or vanishing effective potential. As sketched in the introduction, a sufficient condition for the effective potential $\V$ to be at a (local) minimum, maximum or saddle point is that the configuration does not yield masses $\M$ such that $0<\M<M$. Hence we expect configurations with branes located at $a=0$ and $a={1\over 2}$ to be ``attractors'', as Eq.~(\ref{shift}) shows that in this case  super-Higgs and Higgs effects can combine to yield massless fermions (necessary to have $\nF>0$ as desired), with their bosonic  superpartners acquiring a mass $M$. It turns out that there is another interesting location, $a=\pm {1\over 4}$ (see Fig.~\ref{a1/4}).
The reason why is shown explicitly in Appendix~\ref{A2} but it can be understood qualitatively. 
When strings are stretched between $q$ branes at $a={1\over 4}$ and their mirrors at $a=-{1\over 4}$, they also yield massless fermions. On the other hand, strings with end points at $a=0$ or $\half$ and $a={1\over 4}$ or $-{1\over 4}$ yield bosons and fermions that have ``accidentally'' degenerate masses ${M/2}$. This ``fake supersymmetry'' holds for arbitrary winding number $n_9$ (\ie for the whole KK tower in the original type~I picture). As the leading term in the effective potential is a supertrace, the contributions of these modes to $\V$ cancels. This effect is identical to that described in the heterotic context in Ref.~\cite{CP}.  Eq.~(\ref{Vn}) is still valid  in these special cases, even though as we will see $\V$ has no reason to be generically critical at such a point. 


\subsection{Brane configurations}
\label{BC}

One can deduce general expectations for the dynamical behaviour based on the above formula in  Eq.~(\ref{Vn}), without performing any detailed calculation. Consider for a moment a configuration of D8-branes with $p_1+p_2+2q=32$, as shown schematically on Fig.~\ref{a1/4}.
On geometrical grounds and using Eq.\eqref{shift},  the number of massless bosonic and fermionic degrees of freedom is 
\be
\begin{aligned}
\nB ~=~&\;8\bigg(8+\frac{p_{1}(p_{1}-1)}{2}+\frac{p_{2}(p_{2}-1)}{2}+q\bar{q}\bigg)~, \\
\nF  ~=~&\; 8\bigg(p_{1}p_{2}+\frac{q(q-1)}{2}+\frac{\bar{q}(\bar{q}-1)}{2}\bigg)~,\esp
\label{spectrum}
\end{aligned}
\ee
where $\bar q=q$ (see also Eq.~(\ref{nfbtot})). The reasoning is as follows. In the closed string sector, the massless states surviving the Scherk-Schwarz mechanism are the bosons in the type~IIB theory in 10 dimensions, modded out by the orientifold action. In the NS-NS sector,  the dilaton $\phi$ and metric $G_{MN}$ yield $1+35$ states, while in the RR sector, the 2-form $C_{MN}$ contributes $28$ more states. Altogether, we obtain a contribution of $8\times 8$ degrees of freedom to $\nB$.  
The rest of the massless bosons come from open strings excitations. The gauge group being broken as $SO(32)\rightarrow SO(p_{1})\times SO(p_{2})\times U(q)$, the remaining terms in $\nB$ are the bosonic parts of vector multiplets in the adjoint representations of these groups, while their superpartners are massive due to the Scherk-Schwarz breaking. 
The massless fermionic states {come exclusively from open strings and} are those for which Scherk-Schwarz and WL momentum shifts (in type~I) compensate. Thus, the first term
in $\nF$ comes from massless bifundamental states stretched
between the $p_1$ and $p_2$ $\half$-branes, while the second and third terms come
from states stretched between the branes at $a={1\over 4}$ and their images, filling the antisymmetric and $\overline{\mbox{antisymmetric}}$ representations of $U(q)$. In all these configurations, the effective potential formula~(\ref{Vn}) is valid, and 
\begin{equation}
\label{nbf}
\nF-\nB~=~4\big(\!-(p_{1}-p_{2})^{2}+2(p_{1}+p_{2})-48\big),
\end{equation}
where the RR tadpole cancellation condition (in this case simply $
p_{1}+p_{2}+2q=32)
$ has been used to eliminate $q$. The lowest value $\nF-\nB=-4032$ is reached for $(p_1,p_2,q)=(32,0,0)$ or $(0,32,0)$, and corresponds to a critical point of the potential with respect to the WLs (because $q=0$, implying that no mass scale below $M$ is introduced, as discussed in the introduction).
As a result, we expect that the configuration where the full $SO(32)$ gauge symmetry is restored yields stabilized WLs.\footnote{We say ``expect'' because we have not shown that all critical points of $\V$ satisfy Eq.~(\ref{Vn}). Moreover, we have not shown that all configurations compatible with Eq.~(\ref{Vn}) involve branes at $a=0$, $\half$ and $\pm {1\over 4}$.} The negative potential remains a source for the motion of $M$ (see Refs \cite{CFP,CP,Partouche:2018ftj} for the associated cosmological solutions in flat space). 

The type~I theory admits a second moduli space, disconnected from the one we have just been discussing. In the T-dual picture, this family of models is realized by freezing one $\half$-brane on each O8-plane, implying that the WL matrix can be parameterized as 
\be
\begin{aligned}
\W ~=&~ \diag\big(e^{2i\pi a_\alpha}, \alpha=1,\dots, 32\big)~, \\ 
~\equiv& ~ \diag\big(e^{2i\pi a_1},e^{-2i\pi a_1}, \dots, e^{2i\pi a_{15}},e^{-2i\pi a_{15}},1,-1\big)~,  
\end{aligned}
\label{Wodd}
\ee
with only 15 independent degrees of freedom \cite{Schwarz:1999xj}.  
Restricting as before to configurations with $p_1$ $\half$-branes located at $a=0$, $p_2$ $\half$-branes at $a=\half$, and $q$ branes at $a={1\over 4}$ with their mirrors at $a=-{1\over 4}$, the full gauge symmetry  is $U(1)^2_{G,C}\times SO(p_{1})\times SO(p_{2})\times U(q)$, but now with  $p_1,p_2\in 2\integer +1$. The expression~(\ref{nbf}) is unchanged from earlier, and its minimum $\nF-\nB=-3536$ is reached for $(p_{1},p_{2},q)=(31,1,0)$ or $(1,31,0)$. The associated gauge symmetry is $SO(31)\times SO(1)$, where for notational convenience the inert ``$SO(1)$'' is indicated to remind the presence of an isolated frozen $\half$-brane. As for the $SO(32)$ theory, we expect these configurations to yield stabilized WLs. A crucial observation is that the effective potential for the $SO(31)\times SO(1)$ case, is slightly raised (although still negative) compared to the $SO(32)$ configuration, due to the presence of extra massless fermions and a reduction in $\nB$. This will be of central importance in the next section.


\subsection{Effective potential} 

In order to confirm the above geometrical expectations, let us now present the expression for  the 1-loop effective potential, valid for arbitrary WLs. The calculation is carried out in detail in Appendix~\ref{A2}. In the limit where $M$ is low compared to the string scale,
we find
\be
\V~=~ {\Gamma(5)\over \pi^{14}}\, M^9\sum_{n_9}\frac{\N_{2n_9+1}(\W)}{(2n_9+1)^{10}}~+~{\cal O}\big((M_sM)^{9\over 2}e^{-\pi {M_s\over M}}\big)~,
\ee
where $\N_{2n_9+1}$ is a function that gets contributions from the torus, Klein bottle, annulus and M\"obius strip amplitudes as follows,   
\be
\begin{aligned}
\N_{2n_9+1}(\W)  &~=~4\big(\!-16-0-(\text{tr}\,\W^{2n_{9}+1})^{2}+\text{tr}\,(\W^{2(2n_{9}+1)})\big) \\
&~=~ -16\Bigg(\dis \sum_{\substack{r,s=1\\ r\neq s}}^{N}\cos\!\big(2\pi(2n_9+1)a_{r}\big)\cos\!\big(2\pi(2n_9+1)a_{s}\big)+N-4\Bigg)~,
\end{aligned}
\label{N9}
\ee
where the total number of dynamical $a_r$'s is $N=16$ or 15. Let us analyse this potential for the special cases of ($\half$)-branes located only at $a=0$, $\half$ and $\pm {1\over 4}$~: 

$\bullet$ At such a point in moduli space, $\N_{2n_9+1}$ turns out to be independent of $n_9$,
\be
\N_{2n_9+1} ~=~ \nF-\nB~, 
\ee
and we obtain, as anticipated, 
\be
\V~=~\big(\nF-\nB\big)  \xi_9 M^9~+~\O\big((M_sM)^{9\over 2}e^{-\pi {M_s\over M}}\big)~, 
\ee
where in this case
\be
 \xi_9~=~{\Gamma(5)\over \pi^{14}}\sum_{n_9}\frac{1}{(2n_9+1)^{10}}~=~ \frac{31}{60480\pi^4}~.
\ee

$\bullet$ For $r=1,\dots, N$, the first derivatives are given by  
\be
\begin{aligned}
{\partial \V\over \partial a_r}~=~&\begin{cases} 
~~{\cal O}\big((M_sM)^{9\over 2}e^{-\pi {M_s\over M}}\big)~,&\mbox{ for $a_r=0$ or $\half$}~, \\
~~(p_1-p_2)\xi_9'M^9~+~  {\cal O}\big((M_sM)^{9\over 2}e^{-\pi {M_s\over M}}\big)~,&\mbox{ for $a_r={1\over 4}$}~,\esp
\end{cases} 
\label{dV}
\end{aligned}
\ee
where 
\be
\xi_9' ~=~ {\Gamma(5)\over \pi^{14}}\sum_{n_9\ge 0}\frac{(-1)^{n_9}64\pi}{(2n_9+1)^{9}}~=~ \frac{3}{512\pi^{13}} \left( \zeta(9,{\scriptstyle \frac{1}{4}}) -\zeta(9,{\scriptstyle \frac{3}{4}}) \right)~>~0~,
\ee
and where $\zeta$ is the Hurwitz zeta function. 
Thus, the potential is at a critical point with respect to the WLs only when $q=0$ or $p_1=p_2$, otherwise the branes at $a_r= \frac{1}{4}$ are attracted to the largest  stack located at $a=0$ or $\half$. Notice that for $p_1=p_2$, the brane configuration respects an additional exact symmetry $a \to \half-a$. As a result, we actually expect the generically exponentially suppressed terms in Eq.~(\ref{dV}) to be entirely absent when $p_1=p_2$ (see Footnote~\ref{foo}).

$\bullet$ The  $N\times N$ matrix of second derivatives is block diagonal~:
\be
{\partial^2 \V\over \partial a_ra_s}~=~ \xi_9^{\prime\prime}M^9\bigg(\Big({p_1-p_2 \over 2}-1\Big) I_{\lfloor {p_1/2}\rfloor}, \Big({p_2-p_1 \over 2}-1\Big) I_{\lfloor {p_2/2}\rfloor}, A\bigg)~,
\label{MM}
\ee
where $I_d$ is $d\times d$ identity matrix, $\lfloor x\rfloor$ is the integer part of $x$, $A$ is the $q\times q$ matrix $A_{rs}=\delta_{rs}-1$, and   where
\be
\xi_9^{\prime \prime}~=~{\Gamma(5)\over \pi^{14}}\sum_{n_9}\frac{128\pi^2}{(2n_9+1)^{8}}~.
\ee
{Hence a stable brane configuration must satisfy
\be
{p_1-p_2\over 2}-1\, \ge\,  0\quad \mbox{if}\quad  p_1\, \ge\, 2~ , \qquad \qquad {p_2-p_1\over 2}-1\, \ge\,  0\quad \mbox{if}\quad p_2\,\ge\, 2~,
\ee
whose compatibility implies $p_1$ or $p_2$ to be 0 or 1. When $q=0$, this shows that the $SO(32)$ and $SO(31)\times SO(1)$ configurations are the only stable ones. It turns out that  $q\ge 1$ does not yield other solutions. To see this, note that the vanishing tadpole condition implies $(p_1,p_2,q)=(0,0,16)$ or $(1,1,15)$. 
However, the eigenvalues of $A$ being 1 (with degeneracy $q-1$) and $-(q-1)$ (with degeneracy 1), we conclude that in $U(q)=U(1)\times SU(q)$, even if the $q-1$ WLs of $SU(q)$ are massive, the WL of $U(1)$ is  tachyonic. }

Note that in the quest to find stable (up to exponentially suppressed terms) vacua one might, motivated by Eq.~\eqref{Vn}, have na\"ively looked for  solutions
to $n_{F}^{(0)}-n_{B}^{(0)}=0$. Using Eq.~(\ref{nbf}), this would have yielded theories with open string gauge groups $SO(18)\times SO(14)$, $SO(12)^{2}\times U(4)$ and $SO(14)\times SO(12)\times U(3)$. However, our results show that the first two configurations have tachyonic directions in moduli space, while the third even contains WL tadpoles.


\subsection{Algebraic stability conditions}

The stability conditions of the WLs can also be derived from pure Lie algebra considerations. In quantum field theory, the 1-loop effective potential can be written
as a Schwinger integral (equivalent to the first quantized formalism),
\be
\V~=~-{M_s^9\over 2(2\pi)^9}\int_0^{\infty}{d\tau_2\over \tau_2^{1+{9\over 2}}}\; \mbox{Str}\, e^{-\pi\tau_2\boldsymbol{M}^2}~,
\label{defV}
\ee
where $\boldsymbol{M}$ is the classical mass operator. We are interested in models  where the spectrum arises from massless $\N_{10}=1$ superfields in 10 dimensions, compactified on the Scherk-Schwarz internal circle $S^1(R_9)$. By allowing a WL background, we may have non-trivial $\nF$ and $\nB$ massless states. 
In a Scherk-Schwarzed theory, the existence of full towers of KK modes guarantees that $\V$ is finite even if the domain of integration of the Schwinger parameter contains the UV region $\mbox{$\tau_2\to 0$}$. 

Up to the exponentially suppressed terms arising from string modes heavier than the supersymmetry breaking scale $M$,
the stringy computation yields an identical expression. 
As already mentioned, in such backgrounds, where all modes lighter than $M$ are massless, 
we may switch on WL deformations. Denoting the normalized WL by $y_r$, we can write 
\be
\boldsymbol{M}^2~=~M_s^2\left({m_9+{F\over 2}\over R_9}+Q_ry_r\right)^2~,
\ee
where $Q_r$ is the charge under the $r$-th Cartan $U(1)$ of the gauge group $\G=\prod_\kappa\G_\kappa$. By viewing the supertrace as 
\be
\mbox{Str}~~\equiv~~ \Big(\sum_{{\rm weights}\, Q \in {\cal R}_B}-\sum_{{\rm weights}\, Q\in {{\cal R}_F}}\Big)\sum_{m_9}~,
\ee
where ${{\cal R}_B}, {{\cal R}_F}$ are the representations of the bosonic and fermionic massless states before deformation, Eq.~(\ref{defV}) yields the following at second order in WLs~:
\be
\V~=~M^9\Big(\sum_{{\rm weights}\, Q \in {\cal R}_B}-\sum_{{\rm weights}\, Q\in {{\cal R}_F}}\Big)\Big[ -\xi_9+\#_2\,Q_ry_r\, Q_sy_s+\cdots\Big], \quad \#_2\,>\,0~.
\label{Vexp}
\ee
In this expression, the linear term $\#_1Q_ry_r$ in the brackets is  absent, due to the sum over the weights.\footnote{Moreover, $\#_1$ turns out to vanish, due to the sums over the KK momentum and $F=0,1$. This can be seen after Poisson resummation over $m_9$.} By splitting any representation ${\cal R}$ of $\G$ into a direct product of representations ${\cal R}^{(\kappa)}$ of $\G_\kappa$, we can choose suitable bases of Hermitian generators $T^{(\kappa)}_a$ such that 
\be
\begin{aligned}
T^{(\kappa)}_{\cal R}\delta_{ab}&~=~\half\, \text{tr}\, T^{(\kappa)}_aT^{(\kappa)}_b, \qquad\quad \;\;\;\;\;a,b=1\dots~, \dim \G_\kappa  \\
 \Longrightarrow~~\quad T^{(\kappa)}_{\cal R}\delta_{rs}&~=~\half \sum_{{\rm weights}\, Q \in {\cal R}^{(\kappa)}} Q_rQ_s~, \quad r,s=1,\dots,\mbox{rank}\, \G_\kappa~,\esp
\end{aligned}
\ee
where $T^{(\kappa)}_{\cal R}$ is the Dynkin index of ${\cal R}^{(\kappa)}$. As a result, the squared masses of the WLs $y_r$ are determined by  
\be
{\partial^2\V\over \partial y_r^2}\bigg|_{y=0} ~=~ 4\#_2 M^9 \big(T_{{\cal R}^{(\kappa)}_B}-T_{{\cal R}^{(\kappa)}_F}\big)~.
\ee

In the brane configurations of Sect.~\ref{BC}, we noticed that expression~(\ref{Vexp}) remains true when the undeformed background contains branes located at $\pm{1\over 4}$, thus generating a $U(q)$, $q\ge 1$, gauge group factor, provided $p_1=p_2$. In these circumstances, or when $q=0$, the states charged under $SO(p_1)$ are 8 bosons in the adjoint representation and $8p_2$ fermions in the fundamental. The latter arise from the 8 bifundamentals of $SO(p_1)\times SO(p_2)$. The spectrum charged under $SO(p_2)$ is identical, up to the exchange $p_1\leftrightarrow p_2$. Finally, the states charged under $U(q)$, $q\ge 2$, are 8 bosons in the adjoint, 8 fermions in the antisymmetric and 8 fermions in the $\overline{\mbox{antisymmetric}}$. Table~\ref{DI}, 
\begin{table}
\begin{centering}
\begin{tabular}{|c||c|c|c|c}
\hline 
$\G_\kappa$ & ${\cal R}^{(\kappa)}$ & $\dim {\cal R}^{(\kappa)}$ &  $T_{\cal R}^{(\kappa)}$\\
\hline  \hline 
$SO(p), \; p\ge 2$ & fundamental & $p \esps$ & 1\\
 & adjoint & $\dis \frac{p(p-1)}{2} \esp\espD$ & $p-2$\\
\hline \hline
$SU(q),\; q\ge 2$ & fundamental & $q \esps$& 1\\
 & adjoint & $q^{2}-1 \esp$ & $2q$ \\
  & antisymmetric & $\dis  {q(q-1)\over 2}\esp$  & $q-2$ \\
  & $\overline{\mbox{antisymmetric}} $& $\dis {q(q-1)\over 2} \esp \espD$ & $q-2$ \\
 \hline
\end{tabular}
\par\end{centering}
\caption{Dimensions and Dynkin indexes of representations of simple Lie groups. By convention, the Dynkin index in the fundamental representation is fixed to 1.}
\label{DI}
\end{table}
gives the required Dynkin indexes, from which we find 
\be
{\partial^2\V\over \partial y_r^2}\bigg|_{y=0}~=~64~\#_2M^9~\begin{cases}
\dis {~p_1-p_2\over 2}-1~,\quad \mbox{for the $SO(p_1)$ WLs}\\ \dis {~p_2-p_1\over 2}-1~, \quad \mbox{for the $SO(p_2)$ WLs}\esp\\ ~2~,\;\;\quad \qquad \qquad \mbox{for the $SU(q)$ WLs}~.\esp
\end{cases}
\label{sWL}
\ee
As expected, these results are in agreement with the stability conditions found from the explicit computation of the potential.\footnote{Comparing the eigenvalues of the (mass)$^2$ matrix~(\ref{MM}) with  ${\partial^2\V/ \partial y_r^2}|_{y=0}$, one finds an additional factor of 2 for the $SU(q)$ WLs. This is because contrary to our convention in the algebraic computation, the Dynkin indices of the fundamental representations of $SO(p)$ and $SU(q)$ in the string partition function differ by a factor of 2.}


\section{Non-tachyonic models with \bm $\nF\ge \nB$ in $D$ dimensions at 1-loop}
\label{stabi}

We concluded in the previous section that there are no brane distributions in nine
dimensions that are simultaneously stable with respect to the WLs and yield a non-negative potential. However, between the two stable brane configurations with gauge groups $SO(32)$ and $SO(31)\times SO(1)$, we did note that the latter yields a higher effective potential because of the lower dimension of its gauge group and the presence of extra massless fermions stretching between the two O-planes. This brane setup was stable because the $SO(1)$ factor comes from a frozen $\half$-brane. It seems reasonable to suppose that upon compactification to lower dimensions,
where there are more O-planes in type~II orientifolds, stable configurations might exist in which $\half$-branes are frozen to \emph{different} O-planes, decreasing the dimension of the gauge symmetry and increasing the number of massless fermions even further, and raising the effective potential even more. The hope is that there are then configurations which, apart from being stable with respect to the brane positions, also have $\nF-\nB\ge 0$.


\subsection{Geometric and algebraic picture}
\label{geopic}

To explore this possibility, we will use a  compactification on a torus $T^{10-D}$, with internal metric $G_{IJ}$, $I,J=D,\ldots,9$, and \SS action
always taken to lie in the $9$-th direction. As a result, the scale of supersymmetry breaking is 
\be
M~=~{\sqrt{G^{99}}\over 2}\, M_s~.
\ee 
To specify the open string sector, we choose the ``most geometric picture'' obtained by T-dualizing all of the internal directions, $X^I\to \tilde X^I$. The metric of the dual torus is inverse to the initial one, $\tilde G_{IJ}=G^{IJ}$. The D9-branes and orientifold 9-plane of the type~I theory translate into D$(D-1)$-branes and O$(D-1)$-planes. There is one orientifold plane at each of the $2^{10-D}$ corners of the ``internal box''. 

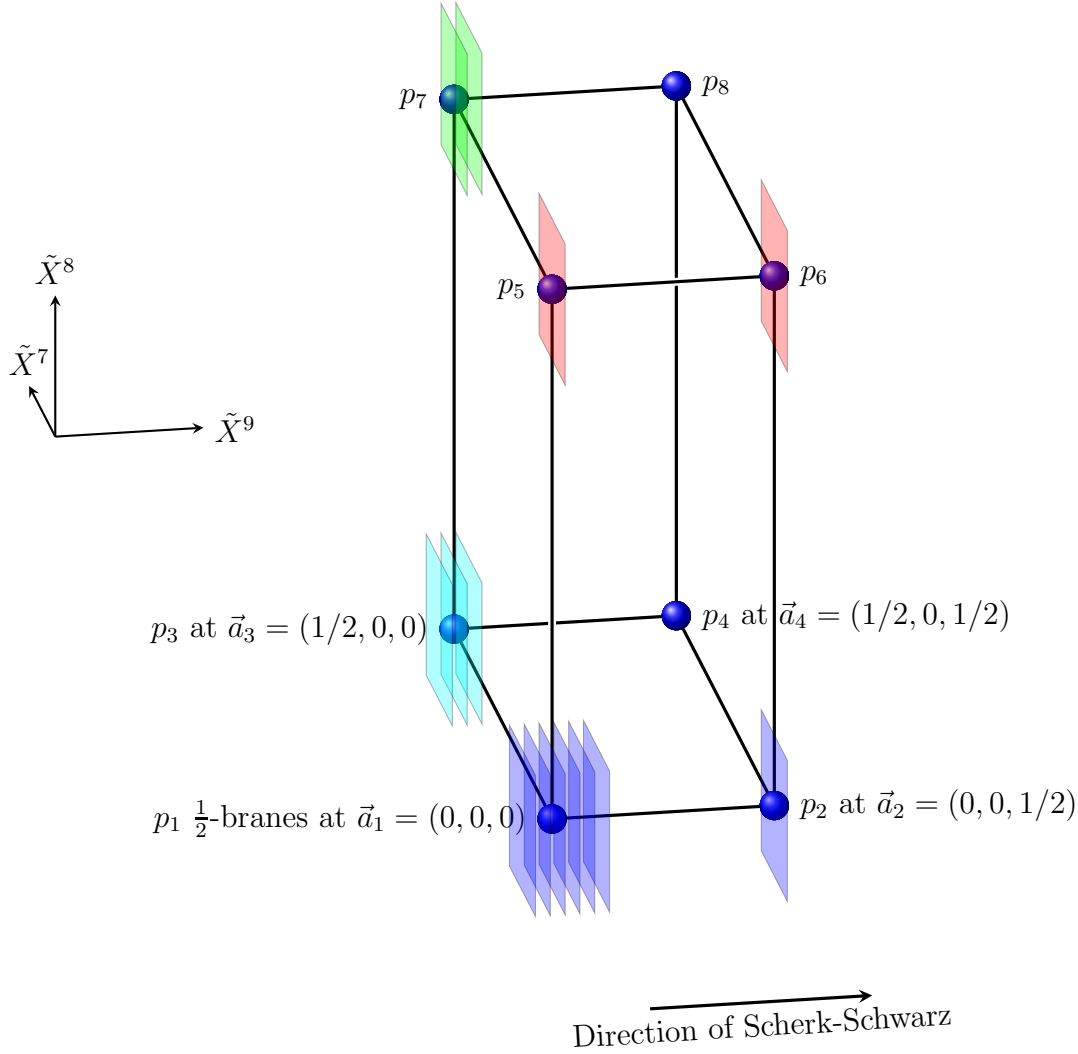
\begin{figure}[h!]
\tikzstyle myBG=[line width=3pt,opacity=1.0]

\tdplotsetmaincoords{70}{80} 

\begin{tikzpicture}[ tdplot_main_coords, >=stealth, cube/.style={very thick,black}, grid/.style={very thin,gray}, axis/.style={->,blue,thick} ] 

\def\scale{3}
\pgfmathsetmacro\wdth{1*\scale} 
\pgfmathsetmacro\hght{2.5*\scale}  
\pgfmathsetmacro\dpth{2.5*\scale} 
\def\branespacing{0.2} 
\pgfmathsetmacro\branescale{\scale/3} 
\def\NoPOne{6} 
\def\NoPTwo{1} 
\def\NoPThree{3} 
\def\NoPFour{0} 
\def\NoPFive{1} 
\def\NoPSix{1} 
\def\NoPSeven{2} 
\def\NoPEight{0} 
\definecolor{ColorPOne}{rgb}{0,0,1} 
\definecolor{ColorPTwo}{rgb}{0,0,1} 
\definecolor{ColorPThree}{rgb}{0,1,1} 
\definecolor{ColorPFour}{rgb}{0,1,1}
\definecolor{ColorPFive}{rgb}{1,0,0} 
\definecolor{ColorPSix}{rgb}{1,0,0} 
\definecolor{ColorPSeven}{rgb}{0,1,0} 
\definecolor{ColorPEight}{rgb}{0,1,0}
\def\BraneOpacity{0.3}
 
\coordinate (p1) at (\dpth,0,0); 
\coordinate (p2) at (\dpth,\wdth,0); 
\coordinate (p3) at (0,0,0); 
\coordinate (p4) at (0,\wdth,0); 
\coordinate (p5) at (\dpth,0,\hght); 
\coordinate (p6) at (\dpth,\wdth,\hght); 
\coordinate (p7) at (0,0,\hght); 
\coordinate (p8) at (0,\wdth,\hght); 

\tikzset{xyp/.style={canvas is xy plane at z=#1}} 
\tikzset{xzp/.style={canvas is xz plane at y=#1}} 
\tikzset{yzp/.style={canvas is yz plane at x=#1}}

\draw[white,myBG]  (p3) -- (p4);
\draw[black,very thick] (p3) -- (p4);
\draw[white,myBG]  (p3) -- (p7);
\draw[black,very thick] (p3) -- (p7);
\draw[white,myBG]  (p4) -- (p8); 
\draw[black,very thick] (p4) -- (p8);
\draw[white,myBG]  (p7) -- (p8);
\draw[black,very thick] (p7) -- (p8);
\draw[white,myBG]  (p1) -- (p5); 
\draw[black,very thick] (p1) -- (p5);
\draw[white,myBG]  (p1) -- (p2);
\draw[black,very thick] (p1) -- (p2);
\draw[white,myBG]  (p2) -- (p6);
\draw[black,very thick] (p2) -- (p6);
\draw[white,myBG]  (p5) -- (p6); 
\draw[black,very thick] (p5) -- (p6);
\draw[white,myBG]  (p3) -- (p1); 
\draw[black,very thick] (p3) -- (p1); 
\draw[white,myBG]  (p4) -- (p2);
\draw[black,very thick] (p4) -- (p2);
\draw[white,myBG]  (p7) -- (p5); 
\draw[black,very thick] (p7) -- (p5);
\draw[white,myBG]  (p8) -- (p6);
\draw[black,very thick] (p8) -- (p6); 

\node at (p3) [circle,fill=black,shading=ball,label=left:{$p_3$ at $\vec{a}_3=(1/2,0,0)$}](ball) {}; 
\node at (p4) [circle,fill=black,shading=ball,label=right:{$p_4$ at $\vec{a}_4=(1/2,0,1/2)$}](ball) {}; 
\node at (p1) [circle,fill=black,shading=ball,label=left:{$p_1$ $\half$-branes at $\vec{a}_1=(0,0,0)$}](ball) {}; 
\node at (p2) [circle,fill=black,shading=ball,label=right:{$p_2$ at $\vec{a}_2=(0,0,1/2)$}](ball) {}; 
\node at (p7) [circle,fill=black,shading=ball,label=left:$p_7$](ball) {}; 
\node at (p8) [circle,fill=black,shading=ball,label=right:$p_8$](ball) {}; 
\node at (p5) [circle,fill=black,shading=ball,label=left:$p_5$](ball) {}; 
\node at (p6) [circle,fill=black,shading=ball,label=right:$p_6$](ball) {};

\tdplotsetrotatedcoords{0}{0}{90} 
\pgfmathsetmacro\NoUpper{floor(\NoPThree/2)-1}
\pgfmathsetmacro\NoLower{-ceil(\NoPThree/2)}
\ifthenelse{\NoPThree=0}{ }{
	\foreach \i in {\NoUpper,...,\NoLower} \fill[opacity=\BraneOpacity,ColorPThree,draw=black,tdplot_rotated_coords,yzp=0.2+\branespacing*\i] (-\branescale,-\branescale) rectangle (\branescale,\branescale);}

\tikzset{shift={(0,\wdth,0)}} 

\pgfmathsetmacro\NoUpper{floor(\NoPFour/2)-1}
\pgfmathsetmacro\NoLower{-ceil(\NoPFour/2)}
\ifthenelse{\NoPFour=0}{ }{
	\foreach \i in {\NoUpper,...,\NoLower} \fill[opacity=\BraneOpacity,ColorPFour,draw=black,tdplot_rotated_coords,yzp=0.2+\branespacing*\i] (-\branescale,-\branescale) rectangle (\branescale,\branescale);}

\tikzset{shift={(\dpth,-\wdth,0)}}

\pgfmathsetmacro\NoUpper{floor(\NoPOne/2)-1}
\pgfmathsetmacro\NoLower{-ceil(\NoPOne/2)}
\ifthenelse{\NoPOne=0}{ }{
\foreach \i in {\NoUpper,...,\NoLower} \fill[opacity=\BraneOpacity,ColorPOne,draw=black,tdplot_rotated_coords,yzp=0.2+\branespacing*\i] (-\branescale,-\branescale) rectangle (\branescale,\branescale);}

\tikzset{shift={(0,\wdth,0)}} 

\pgfmathsetmacro\NoUpper{floor(\NoPTwo/2)-1}
\pgfmathsetmacro\NoLower{-ceil(\NoPTwo/2)}
\ifthenelse{\NoPTwo=0}{ }{
	\foreach \i in {\NoUpper,...,\NoLower} \fill[opacity=\BraneOpacity,ColorPTwo,draw=black,tdplot_rotated_coords,yzp=0.2+\branespacing*\i] (-\branescale,-\branescale) rectangle (\branescale,\branescale);}

\tikzset{shift={(-\dpth,-\wdth,\hght)}} 

\pgfmathsetmacro\NoUpper{floor(\NoPSeven)-2} 
\pgfmathsetmacro\NoLower{-ceil(\NoPSeven/2)}
\ifthenelse{\NoPSeven=0}{ }{
	\foreach \i in {\NoUpper,...,\NoLower} \fill[opacity=\BraneOpacity,ColorPSeven,draw=black,tdplot_rotated_coords,yzp=0.2+\branespacing*\i] (-\branescale,-\branescale) rectangle (\branescale,\branescale);}

\tikzset{shift={(\dpth,0,0)}} 

\pgfmathsetmacro\NoUpper{floor(\NoPFive)-2}
\pgfmathsetmacro\NoLower{-ceil(\NoPFive/2)}
\ifthenelse{\NoPFive=0}{ }{
	\foreach \i in {\NoUpper,...,\NoLower} \fill[opacity=\BraneOpacity,ColorPFive,draw=black,tdplot_rotated_coords,yzp=0.2+\branespacing*\i] (-\branescale,-\branescale) rectangle (\branescale,\branescale);}

\tikzset{shift={(0,\wdth,0)}} 

\pgfmathsetmacro\NoUpper{floor(\NoPSix) - 2}
\pgfmathsetmacro\NoLower{-ceil(\NoPSix/2)}
\ifthenelse{\NoPSix=0}{ }{
	\foreach \i in {\NoUpper,...,\NoLower} \fill[opacity=\BraneOpacity,ColorPSix,draw=black,tdplot_rotated_coords,yzp=0.2+\branespacing*\i] (-\branescale,-\branescale) rectangle (\branescale,\branescale);}

\tikzset{shift={(-\dpth,0,0)}} 

\pgfmathsetmacro\NoUpper{floor(\NoPEight)-2}
\pgfmathsetmacro\NoLower{-ceil(\NoPEight/2)}
\ifthenelse{\NoPEight=0}{ }{
	\foreach \i in {\NoUpper,...,\NoLower} \fill[opacity=\BraneOpacity,ColorPEight,draw=black,tdplot_rotated_coords,yzp=0.2+\branespacing*\i] (-\branescale,-\branescale) rectangle (\branescale,\branescale);}

\tikzset{shift={(2*\dpth,-\wdth,-\hght)}}
\draw[->,very thick,black] (0,0,0) -- (0,\wdth,0) node[midway,sloped,below,align=center] {Direction of Scherk-Schwarz};

\tikzset{shift={(-\wdth,-\dpth,\hght)}} 
\draw[thick,->] (0,0,0) -- (-2,0,0) node[above] {$\tilde X^7$}; 
\draw[thick,->] (0,0,0) -- (0,2,0) node[right] {$\tilde X^9$}; 
\draw[thick,->] (0,0,0) -- (0,0,2) node[above] {$\tilde X^8$}; 
\end{tikzpicture}\caption{\label{dDim}\em \footnotesize Configuration of D7-branes and O7-planes in type~IIB orientifold, in $D=7$ dimensions. At each corner of the internal ``3-box'', there is an orientifold plane coincident with  $p_A$ $\half$-branes, $A=1,\dots,8$. The stacks of $p_{2A-1}$ and $p_{2A}$ $\half$-branes, $A=1\dots, 4$, are separated in  direction $\tilde X^9$, along which the \SS mechanism is implemented. In reality, there are a total of 32 $\half$-branes.}
\end{figure}

Following the route that we took for the nine-dimensional case, it is promising to consider the configurations where the 32 $\half$-branes are coincident with the O$(D-1)$-planes. As explained above, this guarantees that the effective potential satisfies Eq.~(\ref{Vn}) and is critical with respect to the open string WLs. To be specific, we put $p_A$ $\half$-branes on the $A$-th O-plane, $A=1,\dots,2^{10-D}$. In other words, their position along $\tilde X^I$ is $2\pi a_A^I\sqrt{\tilde G_{II}}$, $I=D,\dots,9$, where $a_A^I$ is either 0 or $\half$. Fig.~\ref{dDim} shows the schematic layout of such a brane configuration in 7 dimensions. 
By convention, we order the corners of the box  such that  the $(2A-1)$-th and $2A$-th ones are separated in direction $\tilde X^9$ only, 
\be
a^i_{2A-1}~=~ a^i_{2A}~, \;\;\quad a^9_{2A-1}~=~ a^9_{2A}+\half~,\;\; \quad i\,=\,D,\dots, 8~, \quad A\,=\,1,\dots, 2^{10-D}/2~. 
\ee
The massless spectrum is derived in Appendix~\ref{A3}, Eq.~(\ref{nfbtotD}), but again its counting  can be inferred from geometrical arguments : 
\be
\nB ~=~8\bigg(8+\sum_{A=1}^{2^{10-D}} \frac{p_{A}(p_{A}-1)}{2}\bigg)~,\;\;\quad \nF ~=~8\sum_{A=1}^{2^{10-D}/2}p_{2A-1}p_{2A}~.
\label{nfbtotD'}
\ee
Besides the contribution from the closed string sector,  massless bosons arise from strings attached to a single stack of $\half$-branes, with the bosonic part of vector multiplets arising in the adjoint representation of $SO(p_A)$, $A=1,\dots 2^{10-D}$. On the other hand, massless fermions again occur when the \SS momentum shift in the direction $X^9$ (in the type~I picture) and WL deformations cancel one another. In type~I', this is realized  by strings stretched between two bunches of $\half$-branes separated in the dual  direction $\tilde X^9$.  As a result, they correspond to the fermionic parts of vector multiplets in the bifundamental representation of $SO(p_{2A-1})\times SO(p_{2A})$, $A=1,\dots 2^{10-D}/2$.
Subtracting and using the RR tadpole condition,
$\sum_{A=1}^{2^{10-D}}p_{A}=32$, we obtain
\begin{equation}
\label{nfb0}
\nF-\nB ~=~8\,\bigg(8-\frac{1}{2}\sum_{A=1}^{2^{10-D}/2}(p_{2A-1}-p_{2A})^{2}\bigg)~.
\end{equation}

Next, the generic algebraic derivation of the WLs stability condition in nine dimensions can also be generalised.  Denoting again the gauge symmetry group as $\G=\prod_\kappa\G_\kappa$, the $r$-th Cartan $U(1)$, $r=1,\dots, \mbox{rank}\,  \G$, admits  WLs  denoted $y_{r}^I$ along the internal directions $I=D,\dots,9$.\footnote{As in 9 dimensions, we assume here that in the undeformed background, all states lighter than $M$ are massless.} Taylor expanding the potential, one obtains \cite{CP,SNSM,SNSM2} 
\be
\V~=~M^D\Big(\sum_{{\rm weights}\, Q \in {\cal R}_B}-\sum_{{\rm weights}\, Q\in {{\cal R}_F}}\Big)\Big[ -\xi_D+\#_2\,Q_rQ_s\Big(\sum_{i=D}^8{y_{r}^iy_{s}^i\over (D-1)G^{99}}+y_{r}^9 y_{s}^9\Big)+\cdots\Big]~,
\label{VexpD}
\ee
where $\#_2>0$ and where  ${\cal R}_B$, ${\cal R}_F$ are the representations of the massless bosons and fermions at the critical point. As a result, the (in)stability of the WLs $y_{r}^I$ associated to the gauge group factor $\G_\kappa$ is independent of the choice of Cartan $U(1)\subset \G_\kappa$, and of the direction $I$. Applying this rule to our case of interest, we have $SO(p_{2A-1})$ and  $SO(p_{2A})$ gauge group factors, with $8$ bosons in their adjoint representations,  and respectively $8p_{2A}$ and $8p_{2A-1}$ fermions in their fundamental representations (arising from the bifundamentals). Thus, the conditions for the WLs not to be tachyonic at 1-loop are, for $A=1,\dots, 2^{10-D}/2$, 
\be
\begin{cases}
~p_{2A-1}-2-p_{2A}~\ge~ 0~, \quad \mbox{for the $SO(p_{2A-1})$ WLs~,\;\;\; if $p_{2A-1}~\ge~ 2~,$}\\ 
~p_{2A}-2-p_{2A-1}~\ge~ 0~, \quad \mbox{for the $SO(p_{2A})$ WLs~,\;\;\;\;\quad  if $p_{2A}~\ge~ 2$}~,\esp\esp
\end{cases}
\label{mas}
\ee
where we have used the Dynkin indices of Table~\ref{DI}. Compatibility of these constraints forces either $p_{2A-1}$ or $p_{2A}$ to be 0 or 1. The conditions are even more severe when either $p_{2A-1}$ or $p_{2A}$ equals 2, in which case the only allowed choices are  $(p_{2A-1},p_{2A})=(2,0)$ or $(0,2)$. Finally, $p_{2A-1}$ and $p_{2A}$ both equal to 0 or 1 is also trivially possible, since there are no WLs associated to the ``gauge group factors'' $SO(1)$ and $SO(1)\times SO(1)$.

Returning to our specific type~II orientifold setup,  when all $\half$-branes are located at the corners of the internal box, the effective potential is critical,   and we are looking for the configurations satisfying  
\begin{align}
\label{const}
&(i)\;\;\;\; \sum_{A=1}^{2^{10-D}}p_{A}~=~32,\qquad \qquad \qquad\qquad \qquad \qquad   \qquad \qquad \quad\qquad \; \mbox{(RR tadpole cancellation)} \nonumber \\
&(ii)\;\;\sum_{A=1}^{2^{10-D}/2}(p_{2A-1}-p_{2A})^{2}~\le~16, \qquad \qquad \qquad\qquad \qquad \qquad  \qquad  \mbox{($\nF-\nB\, \ge\, 0$)}\\
&(iii)\;\;\;\;\forall A=1,\dots, 2^{10-D}/2,\;\;\; (p_{2A-1},p_{2A})= \begin{cases} 
\phantom{\mbox{or }}(p,0),  (0,p),\;\; p\ge 0\quad\;\;\mbox{(No tachyonic open}\\ 
\mbox{or }(p,1),(1,p),\;\; p\ge 3 \mbox{ or } p=1\quad\,  \mbox{string WL).}
\end{cases} \nonumber 
\end{align}
These conditions admit many solutions. For instance, Table~\ref{models} displays all corresponding gauge groups generated by the open strings, when $\nF-\nB=0$. Their rank varies between 4 and 0. We denote by $[SO(p)\times SO(1)]$ the gauge symmetry realized by stacks of $\half$-branes such that $(p_{2A-1}, p_{2A})=(p,1)$ or  $(1,p)$. Similarly, $SO(p)$ is the gauge group factor generated by stacks where $(p_{2A-1}, p_{2A})=(p,0)$ or  $(0,p)$. 
\begin{table}[h]
\begin{centering}
\begin{tabular}{|l|c|}
\hline 
open string gauge group & $D\le$   \tabularnewline
\hline 
\hline 
$\big[SO(5)\times SO(1)\big]\times\big[SO(1)\times SO(1)\big]^{13}$ & 5 \tabularnewline
\hline 
$SO(4)\times\big[SO(1)\times SO(1)\big]^{14}$ & 5\tabularnewline
\hline 
$\big[SO(4)\times SO(1)\big]\times\big[SO(3)\times SO(1)\big] \times SO(1)^3\times\big[SO(1)\times SO(1)\big]^{10}$ &5  \tabularnewline
\hline 
$\big[SO(4)\times SO(1)\big]\times SO(2) \times SO(1)^3\times\big[SO(1)\times SO(1)\big]^{11}$ & 5 \tabularnewline
\hline 
$\big[SO(4)\times SO(1)\big]\times SO(1)^7\times\big[SO(1)\times SO(1)\big]^{10}$ & 4 \tabularnewline
\hline $SO(3)\times\big[SO(3)\times SO(1)\big] \times SO(1)^3\times\big[SO(1)\times SO(1)\big]^{11}$ & 5 \tabularnewline
\hline
 $SO(3)\times SO(2) \times SO(1)^3\times\big[SO(1)\times SO(1)\big]^{12}$ &4  \tabularnewline
\hline
$SO(3)\times SO(1)^7\times\big[SO(1)\times SO(1)\big]^{11}$ &  4\tabularnewline
\hline
$SO(2)^u\times \big[SO(3)\times SO(1)\big]^v\times SO(1)^{16-4(u+v)}\times \big[SO(1)\times SO(1)\big]^{8+u},u+v\le 4$ &  \tabularnewline
\hline 
\end{tabular}
\par\end{centering}
\caption{\em \footnotesize Gauge symmetry groups realized by open strings,  in models where the positions of all $\half$-branes (in type~II orientifolds) are at corners of the ``internal box'' and (marginally) stable, when $\nF-\nB=0$. An $[SO(p)\times SO(1)]$ factor arises from a stack of $p$ $\half$-branes and a single $\half$-brane  located at corners separated along the \SS direction $\tilde X^9$ only. An $SO(p)$ factor is realized by a  stack of $p$ $\half$-branes at some corner, with no other $\half$-brane at the corner separated along $\tilde X^9$. As indicated in the last column, maximal spacetime dimensions are required for these configurations to exist. In the last line, the gauge groups can be realized in dimension $D\le 5$ when $u+v=4$ or $(u,v)=(0,3)$, $(1,2)$, while all other values of $(u,v)$ require $D\le 4$.}
\label{models}
\end{table}
Note that even though exchanging $p_{2A-1}\leftrightarrow p_{2A}$ or  $(p_{2A-1},p_{2A})\leftrightarrow (p_{2B-1},p_{2B})$ for some $A,B=1,\dots, 2^{10-D}/2$ does not change the massless spectrum, the  models are not equivalent in general. This is due to the massive open strings stretched between bunches of $\half$-branes separated in some direction(s) $\tilde X^i$, $i=D,\dots, 8$. Thus, it is understood that each line in Table~\ref{models} corresponds to a class of models obtained by inequivalent permutations of the stacks of $\half$-branes. For a gauge group to be realized in dimension $D$, the total number of $[SO(p)\times SO(1)]$ and $SO(p)$ factors must be lower or equal to the number of pairs of corners, $2^{10-D}/2$. As a result, there is a maximal spacetime dimension for each brane configuration, as indicated in the second column of Table~\ref{models}. In the last line, $D\le 5$ is required when  $u+v=4$ or $(u,v)=(0,3)$, $(1,2)$, while all other values of  $(u,v)$ are allowed when   $D\le 4$.  For completeness, we note  that the closed string sector also contributes Abelian factors  $U(1)^{2(10-D)}_{G,C}$, which arise from the reductions of the metric and RR 2-form, $G_{\mu I}$, $C_{\mu I}$, $I=D,\dots, 9$. 
The constraints~(\ref{const}) turn out to be more severe as $\nF-\nB$ increases. In fact, both the number of allowed configurations and the maximal value the $p_A$'s can take decrease. In particular,  the highest value $\nF-\nB=8\times 8$ is obtained for a unique configuration $[SO(1)\times SO(1)]^{16}$, in dimension $D\le 5$. In this case, the gauge symmetry arises from the closed string sector, $U(1)^{10-D}_G \times U(1)^{10-D}_C$, while the open strings provide neutral fermions.    

So far we have discussed configurations in which the open string WLs are not tachyonic. However, marginal stability of the brane system is not guaranteed in the special case that  some of the WLs are massless, since higher order interactions (still at 1-loop) may introduce instabilities. Therefore, we have to analyze in more detail the WL deformations of group factors $SO(2)$ and $[SO(3)\times SO(1)]$, which are of rank 1 and yield massless WLs. The case of the $SO(2)$ factors is easy to treat, because the  light spectrum of the theory is neutral under them. Hence, the masses of these light  states  do not depend on the $SO(2)$ WLs along $T^{10-D}$ (as there is no possible Higgs mechanism). As a result, the effective potential defined in Eq.~(\ref{Vdef}), which depends only on the mass spectrum, has a dominant contribution trivially independent of these WLs. In other words, the latter are  flat directions, up to exponentially suppressed terms. On the contrary, $[SO(3)\times SO(1)]$ WLs do modify the masses of the non-Cartan bosonic states in the adjoint representation, as well as those of the fermions in the fundamental. Thus, it is only by careful study of the effective potential in the next subsection that we will be able to conclude that they also do not induce instabilities.

Finally, we should mention that the reader interested only in stable brane configurations irrespective of the sign of $\nF-\nB$ can simply relax constraint~($ii$) in Eq.~(\ref{const}). In that case, the allowed open string gauge groups are 
\be
\prod^P_{\rho=1} SO(p_\rho)\prod^Q_{\omega=1}[SO(p'_{\omega})\times SO(1)]\,,\quad \where \quad \sum^P_{\rho=1} p_\rho +\sum^Q_{\omega=1}p'_{\omega}+Q \,=\,32\,, \quad P+ Q \,\le\, 2^{10-D}/2~.
\ee
The number of solutions is drastically increased, as are the ranks and the dimensions of the groups. For instance,  the $SO(32)$ brane configuration is unsurprisingly stable  in arbitrary dimension. 


\subsection{Effective potential}
\label{vefD}

The conclusions made above on geometric and algebraic grounds can of course be recovered directly from the 1-loop effective potential, which is derived for any spacetime dimension in Appendix~\ref{A3}. A first way to write the result combines the torus, annulus and M\"obius strip amplitudes given in Eqs~(\ref{TD}), (\ref{AD}), (\ref{MD}), while that of the Klein bottle vanishes. These expressions being valid at  arbitrary closed and open string moduli, they necessarily incorporate the whole spectrum of the theory, the notion of light (and thus dominant) modes being location-dependent in moduli space. Therefore, it is more illuminating to specify an initial background and consider fluctuations in a local neighbourhood. 

Our choice of background is as described in the previous subsection, with $p_A$ $\half$-branes coincident at the $A$-th corner with an orientifold plane, $A=1,\dots,2^{10-D}$. The only constraint we impose here is  RR tadpole cancellation, \ie $\sum_{A=1}^{2^{10-D}}p_A=32$. We will denote by $a_\alpha^I$ the WL associated to the $\alpha$-th $\half$-brane along the direction $I$. For the $p_A$ $\half$-branes situated in the vicinity of corner $A$, we may then write  
\be
a_\alpha^I~=~a_A^I+\varepsilon_\alpha^I~, \quad\;\; \where\;\; \quad a_A^I\in\Big\{0, \half\Big\}~,\;\; \quad I\,=\,D,\dots, 9~.
\ee
Notice that the $\varepsilon_\alpha^I$'s are not all dynamical degrees of freedom. In fact, $\varepsilon_\alpha^I\equiv -\varepsilon_\beta^I$ when $\alpha$ and $\beta$ label branes images of one another. Moreover, if $p_A$ is odd, in addition to the pairs of such $\half$-branes, there is always one left over, say the $\alpha$-th, which is frozen at the corner, $\varepsilon^I_\alpha\equiv 0$.

In this subsection we will also choose the internal metric $G_{IJ}$, $I,J=D,\dots,9$,  so that all KK and winding mass scales are greater than the supersymmetry breaking scale. We also take the latter to be lower than the string scale, in order to avoid any Hagedorn-like tree-level instability. In total then, we assume
\be
G^{99}\,\ll \, | G_{ij}|  \,\ll\, G_{99}~, \quad |G_{9j}|\,\ll\,  \sqrt{G_{99}}~,\quad i,j\,=\,D,\dots,8~ ,\quad G_{99}\gg 1~.
\label{hyp}
\ee 
In Fig.~\ref{dDim}, this is shown in the type~II orientifold picture by the fact that  the length of the T-dual direction $\tilde X^9$ is smaller than those of the transverse directions $\tilde X^i$, $i=D,\dots, 8$.   
The remaining closed string moduli are the internal RR 2-form components,  $C_{IJ}$. They  combine with the Abelian vector fields $C_{\mu J}$ to make up the  bosonic parts of vector  multiplets of the underlying spontaneously broken maximally supersymmetric  theory. As a result, they can be interpreted as WLs along $T^{10-D}$ associated to the gauge group $U(1)^{10-D}_C$. However, there is no state in the perturbative spectrum, charged under these $U(1)$'s. As a result, the tree level mass spectrum cannot depend on the  marginal deformations $C_{IJ}$ of the worldsheet CFT, and the same is  true for the 1-loop potential~(\ref{Vdef}). Hence, the marginal stability of the RR moduli $C_{IJ}$ is preserved at the 1-loop level.  

Under the above assumptions, we find 
\be
\V~=~   {\Gamma\big({D+1\over 2}\big)\over \pi^{3D+1\over 2}}\, M^{D}\sum_{l_9}{\hat \N_{2l_9+1}(\varepsilon,G)\over |2l_9+1|^{D+1}} ~+~\O\big((M_sM)^{D\over 2}e^{-2\pi c {M_s\over M}}\big)~,
\label{vtot}
\ee
where $c=\O(1)$ is positive, and  
where $\hat \N_{2l_9+1}$ combines the torus amplitude, Eqs~(\ref{TDs}), the trivial Klein bottle contribution, as well as the annulus and M\"obius strip contributions, Eqs~(\ref{ADs}), (\ref{MDs})~:
\begin{align}
\label{hatn}
\hat \N_{2l_9+1}(\varepsilon,G)~ = &~4\,\bigg\{\!-16-0-\sum_{(\alpha,\beta)\in L}(-1)^{F}\cos\!\Big[2\pi(2l_9+1)\big(\varepsilon_\alpha^9-\varepsilon_\beta^9+{G^{9i}\over G^{99}}(\varepsilon_\alpha^i-\varepsilon_\beta^i)\big)\Big]\nonumber \\
 &  \qquad\qquad\qquad\qquad  \times ~\Hc_{D+1\over 2}\bigg(\pi|2l_9+1|{\big[(\varepsilon_\alpha^i-\varepsilon_\beta^i)\hat G^{ij}(\varepsilon_\alpha^j-\varepsilon_\beta^j)\big]^\half\over \sqrt{G^{99}}}\bigg)\\
& ~+ ~\sum_\alpha  \cos\!\Big[4\pi(2l_9+1)\big(\varepsilon_\alpha^9+{G^{9i}\over G^{99}}\, \varepsilon_\alpha^i\big)\Big]  \Hc_{D+1\over 2}\bigg(2\pi|2l_9+1|{\big[\varepsilon_\alpha^i\,\hat G^{ij}\,\varepsilon_\alpha^j\big]^\half\over \sqrt{G^{99}}}\bigg)\bigg\}~.\nonumber \esp
\end{align}
In this expression, $L$ is the set of pairs $(\alpha,\beta)$ such that $\alpha$ and $\beta$ are $\half$-branes in the neighbourhood of {either of the corners $2A-1$ and $2A$,} for some $A=1,\dots,2^{10-D}/2$. The sectors $(\alpha,\beta)$ yield light strings stretched between these $\half$-branes that  generate the bosonic adjoint and fermionic bifundamental representations of $SO(p_{2A-1})\times SO(p_{2A})$.  In our notation, $F$ is the fermionic number of these modes. Moreover, we have defined an effective inverse metric 
\be
\hat G^{ij}~=~ G^{ij}-{G^{i9}\over G^{99}}\, G^{99}\, {G^{9j}\over G^{99}} ~=~ G^{ij}+\O\Big({1\over G^{99}}\Big)~, \quad i,j=D,\dots, 8~,
\label{hG2}
\ee
for the internal space transverse to the larger \SS direction, $X^9$. The index $l_9$ is obtained  by Poisson resummation over the KK momentum $m_9$ (of the initial type~I picture) along $X^9$. Finally, the function $\Hc_\nu$ is defined in Eq.~(\ref{dexpsup}). {From} this result, it is natural to parametrise the NS-NS moduli space by $(\hat G^{ij}, G^{9i}, G^{99})$. Some further remarks are in order~: 

$\bullet$ In the initial background, where all $\half$-branes are located at corners, we have $\varepsilon_\alpha^I=0$, $\alpha=1,\dots,32$,  $I=D,\dots,9$, implying that $\hat \N_{2l_9+1}$ becomes $l_9$-independent there, 
\be
\hat \N_{2l_9+1}(0,G) ~=~ \nF-\nB~. 
\ee
Hence, as expected, the effective potential satisfies the rule-of-thumb,
\be
\V~=~\big(\nF-\nB\big)  \xi_D  M^D~+~\O\big((M_sM)^{D\over 2}e^{-2\pi c {M_s\over M}}\big)~, \quad \where \quad  \xi_D~=~{\Gamma({D+1\over 2})\over \pi^{3D+1\over 2}}\sum_{n_9}\frac{1}{|2n_9+1|^{D+1}}~,
\label{rule}
\ee
where the pure KK modes dominate, and all other states at mass scales greater than $M$ yield exponentially suppressed contributions.

$\bullet$ Applied to backgrounds satisfying the RR tadpole and stability constraints $(i)$ and $(iii)$ in Eq~(\ref{const}), we have argued that the expression of the potential yields non-tachyonic  open string WLs at 1-loop. To check this, let us denote in this paragraph the dynamical WL degrees of freedom as
\be
\varepsilon_r^I~, \quad I=D,\dots, 9~, \quad r=1,\dots ,  \sum_{A=1}^{2^{10-D}} \left\lfloor {p_A\over 2}\right\rfloor~,
\ee
and Taylor expand $\hat \N_{2l_9+1}(\varepsilon,G)$ to quadratic order in $\varepsilon_r^I$ :
\be
\begin{aligned}
\V&~=~\big(\nF-\nB\big)  \xi_D  M^D ~+~ \half\,  \xi_D^{\prime\prime}M^D\sum_r\left({p_{A(r)}-p_{\tilde A(r)}\over 2}-1\right)\varepsilon_r^I\hat \Delta_{IJ}\varepsilon_r^J\\
&\hspace{7cm}~+~\O(\varepsilon^4)
~+~\O\big((M_sM)^{D\over 2}e^{-2\pi c {M_s\over M}}\big)~.
\end{aligned}
\ee
In the above expression, we have defined
\be
\xi_D^{\prime \prime}~=~{\Gamma({D+1\over 2})\over \pi^{3D+1\over 2}}\sum_{n_9}\frac{128\pi^2}{|2n_9+1|^{D-1}}~, \qquad \hat \Delta_{IJ}~=~{1\over D-1}\!\left({G^{IJ}\over G^{99}}+(D-2){G^{I9}\over G^{99}}{G^{9J}\over G^{99}}\right)~.
\ee
Moreover, $A(r)$ denotes the corner around which the brane $r$ varies,  while  $\tilde A(r)$ is the partner corner along the \SS direction $\tilde X^9$.  The entries in the mass matrix are of the form 
\be
\hat \Delta_{IJ}~=~ \begin{pmatrix}{G^{ij}\over (D-1)G^{99}}+\O(1)& \O(1)\\ \O(1)& 1\end{pmatrix}~,
\ee
from which we conclude that $\hat \Delta$ has positive eigenvalues:  $9-D$ of them are large, $\O(G_{99})$, while the last one is $\O(1)$. This result is in  perfect agreement with the field theoretic expectation, Eq.~(\ref{VexpD}), and the heterotic result \cite{CP}. Hence, the stability (or flatness at 1-loop) conditions Eq.~(\ref{mas}) are recovered. 

$\bullet$ When stacks of branes satisfy $(p_{2A-1},p_{2A})=(2,0)$, $(0,2)$, $(3,1)$, $(1,3)$, the dynamical WLs are massless. For $(p_{2A-1},p_{2A})=(2,0)$ or $(0,2)$, there is in the vicinity of one corner a single brane at position $\varepsilon^I_\alpha$, and its mirror at $\varepsilon^I_\beta\equiv -\varepsilon^I_\alpha$. Extracting the contributions of these branes from $\hat\N_{2l_9+1}$, we find that they induce a term in $\V$ of the form  
\[
{\Gamma\big({D+1\over 2}\big)\over \pi^{3D+1\over 2}}\, M^{D}\sum_{l_9}{-8\over |2l_9+1|^{D+1}},
\]
which is independent of the degrees of freedom $\varepsilon_\alpha^I$. In fact, all cosines and $\Hc$ functions with non-trivial $\varepsilon_\alpha^I$-dependance cancel one another, as expected from our previous arguments. We can proceed the same way for $(p_{2A-1},p_{2A})=(3,1)$ or $(1,3)$. In these cases, there is one dynamical brane, its mirror and one frozen $\half$-brane at one corner, and another frozen $\half$-brane at the second corner. Extracting from $\hat \N_{2l_9+1}$ all the contributions arising from them, the result turns out to vanish identically (!)  In other words, the dominant contribution of $\V$ does not depend on the degrees of freedom associated to such  subsystems of branes. To summarize, the tree level marginal stability of the WLs associated to $SO(2)$ and $[SO(3)\times SO(1)]$ gauge factors remains valid at 1-loop, up to exponentially suppressed terms.

$\bullet$ When condition~($iii$) in Eq.~(\ref{const}) is satisfied, we conclude that (keeping  $M$ fixed) $\V$ is at a local minimum when all {\em massive} WL fluctuations $\varepsilon_\alpha^I$ are set to 0. Since this is irrespective of the values of the massless open string WLs (as well as those arising from the RR sector, $C_{IJ}$), Eq.~(\ref{rule}) is valid in this more general case. It is then clear that the NS-NS moduli  $\hat G^{ij}$, $G^{9i}$, $i,j=D,\dots, 8$, are additional flat directions of these minima (up to the exponentially suppressed terms). Hence, the only non (marginally) stabilized modulus at the 1-loop level is $M= M_s \sqrt{G^{99}}$, unless $\nF-\nB=0$.

To summarize, we have found local minima of arbitrary signs of the open string effective potential, at fixed $M$, up to exponentially suppressed terms, and valid at 1-loop.  These minima are degenerate, with flat directions parametrized by the massless open string WLs, the RR moduli and all NS-NS moduli fields except $M$, unless the minimum vanishes. Note that from  the adiabatic argument of Ref.~\cite{Vafa:1995gm} applied to the heterotic / type I duality (see Sect.~\ref{sec.4}),  one expects that some  closed string moduli may be stabilized, though nonperturbatively from the type~I point of view~\cite{Estes:2011iw}.
Finally, let us stress that the above results were derived assuming Eq.~(\ref{hyp}) is satisfied, which guarantees that ultimately the NS-NS metric components live in a (very) large plateau of the effective potential. The goal of the next section is to see which configurations remain (marginally) stable at 1-loop if one ventures outside the region defined  by Eq.~(\ref{hyp}). 


\subsection{Extension of the domain of validity}
\label{vefD2}

For the backgrounds presented so far to be marginally stable at 1-loop, we have imposed that $G_{ij}$, $G_{9j}^2$ are bounded from above by $G_{99}\gg 1$. We would now like to relax this condition, at the expense of further restricting the brane configurations. To be concrete, let us consider the region of moduli space in which all internal directions of the original type~I picture are bigger than the string scale, \ie
\be
G_{II}~\gg~ 1, \qquad \mbox{(no sum on) }I\,=\,D,\dots,9~,
\label{newhyp}
\ee 
This region in moduli space covers partially that considered in Eq.~(\ref{hyp}), but also 
allows $G_{ii}$, for some $i=D,\dots,8$, to be greater than $G_{99}$. The backgrounds marginally stable at 1-loop in both regions~(\ref{newhyp}) and~(\ref{hyp})  will be a subset of those specified in Sect.~\ref{vefD}. 

Under the above assumption, the 1-loop potential, which is computed at the end of Appendix~\ref{A3}, takes the form
\be
\V~=~{\Gamma(5)\over \pi^{D+5}}\, {M_s^D\over 2^D} \sqrt{\det G}  \sum_{\vec l}{\N_{\tilde l}(\vec \W)\over (\tilde l_IG_{IJ}\tilde l_J)^5}~+~\O\big(M_s^D\sqrt{\det G}\, G_{99}^{-{11\over 4}}e^{-2\pi\sqrt{G_{99}}}\big)~.
\label{VTOT11}
\ee
where we use the notation $\vec \W\equiv (\W_D,\dots,\W_9)$ to encode arbitrary open string WL matrices,
\be
\W_I~=~ \diag\!\Big(e^{2i\pi a^I_\alpha}, \alpha=1,\dots,32\Big),\quad I\,=\,D,\dots,9~, 
\label{WI2}
\ee
and where $\tilde l\equiv (l_D,\dots,l_8,2l_9+1)$ is a vector whose last  integer entry (associated to the Scherk-Schwarz direction) is odd. The numerator  
\be
\N_{\tilde l}(\vec \W)~=~4\Big(\!-16-0-\big(\text{tr}\, (\W_D^{\tilde l_D}\cdots\W_9^{\tilde l_9})\big)^{2}+\text{tr}\, (\W_D^{2\tilde l_D}\cdots\W_9^{2\tilde l_9}\big)\Big)~,
\label{VTOT}
\ee
contains four contributions respectively arising from the torus, Klein bottle, annulus and M\"obius strip amplitudes. 

To understand this result, first consider all $\half$-branes to be coincident with the O$(D-1)$-planes, $p_A$ of them sitting at the $A$-th corner, at  position parameterized by $\vec a_A\equiv (a_A^D,\cdots, a_A^9)$, $a_A^I\in\{0,\half\}$, $A=1,\dots, 2^{10-D}$, $I=D,\dots,9$. In that case, we have
\be
\N_{\tilde l}(\vec \W)~=~8\bigg(-8-\half\sum_{A,B=1}^{2^{10-D}/2} (p_{2A-1}-p_{2A})(p_{2B-1}-p_{2B}) (-1)^{2\vec l'\cdot (\vec a_A'-\vec a_B')}+\frac{1}{2}\sum_{A=1}^{2^{10-D}}p_{A}\bigg)~,
\ee
where all ``primed'' vectors have entries $i=D,\cdots, 8$ only, \ie  $\vec V\in \Z^{10-D} \Rightarrow \vec V\equiv (\vec V',0)$. Notice that all terms $A\neq B$ are dressed with an alternative $\vec l'$-dependent sign, while all other contributions count the massless bosons and fermions of the configuration, Eq.~(\ref{spe}). In fact, taking the internal metric $G_{IJ}$ to satisfy Eqs~(\ref{hyp}), the contributions $A\neq B$ are exponentially suppressed, letting the remaining terms reproduce Eq.~(\ref{rule}). Physically, this is clear since strings stretched between corners $2A-1$ or $2A$ at one end, and $2B-1$ or $2B$ at the other end, become super heavy compared to the supersymmetry breaking scale $M$, when $A\neq B$. However, when some of these strings become lighter than $M$, which is allowed by Eq.~(\ref{newhyp}), their contributions are no longer negligible and appear in Eq.~(\ref{VTOT}) with sector-dependent dressing functions
\[
 \sum_{\vec l}{(-1)^{2\vec l'\cdot (\vec a_A'-\vec a_B')}\over (\tilde l_IG_{IJ}\tilde l_J)^5}~, \qquad A\neq B\,=\,1,\dots,2^{10-D}/2~.
\]
As a result, the assumption used in the previous sections that all mass scales below $M$ vanish in the undeformed backgrounds is no longer valid and our algebraic derivation of the stability conditions do not apply. What we see explicitly here is that when the above sector-dependent functions are present, finding a marginally stable point in moduli space may be hard, if not impossible, at least for the internal metric components. 

 
 The interpretation of such dressing functions is that the exponential terms in the potential that we have so far been neglecting may become large when $G_{ii} \gg G_{99}$ (\ie $R_i \gg R_9$ in an untilted torus). Indeed 
on an untilted torus such terms are best evaluated  by Poisson resumming direction $9$ only and making a saddle point approximation, which yields a contribution proportional to $e^{-2\pi \sqrt{G_{99}/G_{ii}}} = e^{-2\pi R_9/R_i}$. The physical meaning of such factors, which can be important when $R_i > R_9$, is that the KK modes in the $i$-th direction (with masses going like $1/R_i$ in the type~I setup) have to traverse the entire Scherk-Schwarz direction 9 before they can feel the supersymmetry breaking, so they contribute to the potential with the typical Yukawa factor. As $R_i$  increases in size the KK modes become light enough that this is no longer a suppression, and the contribution can no longer be neglected. 

The rule-of-thumb then is that a direction is allowed to become large (in the original type~I picture) as long as the Scherk-Schwarz breaking is Bose-Fermi degenerate or absent for its KK modes. This is equivalent to considering brane configurations such that $\N_{\tilde l}(\vec \W)$ is $\vec l'$-independent, which is the case when all but one pair of corners $(2B-1,2B)$ satisfy $p_{2B-1}=p_{2B}$. Up to a relabelling, we will take the remaining couple of corners to be $A=1$ and 2. In fact, for such backgrounds to be marginally stable at 1-loop at least in region~(\ref{hyp}) of the moduli space, we must also impose condition $(iii)$ in Eq.~(\ref{const}) : 
\be
\begin{aligned}
 p_{2B-1}& ~=~p_{2B}\in\{0,1\}, \quad B\,=\,2,\dots,2^{10-D}/2~,\\
(p_1,p_2)& ~\in~\{(2p,0), (2p-1,1)\},\quad\where\quad p+ \sum_{B=1}^{2^{10-D}/2}p_{2B-1}~=~16~.
\end{aligned}
\ee
In this case, the dynamical open string WLs are those associated to  $SO(p_1)$, and the corresponding degrees of freedom can be defined as 
\be
(a_\alpha^I-a_{A=1}^I, \alpha=1,\dots, p_1)~=~\begin{cases}\Big(\varepsilon_1^I, -\varepsilon_1^I,\dots, \varepsilon_{p_1\over2}^I,-\varepsilon_{p_1\over 2}^I\Big)\qquad \;\,\quad \mbox{for $p_1$ even}\espD\\
\Big(\varepsilon_1^I, -\varepsilon_1^I,\dots, \varepsilon_{p_1-1\over2}^I,-\varepsilon_{p_1-1\over 2}^I,0\Big)\quad \mbox{for $p_1$ odd}~.
\end{cases}
\ee
In these variables, we obtain
\be
\N_{\tilde l}(\vec \W)~=~ -16\Bigg(\dis \sum_{\substack{r,s=1\\ r\neq s}}^{\half(p_1-p_2)}\cos(2\pi \tilde l \cdot \vec \varepsilon_r)\cos(2\pi \tilde l \cdot \vec \varepsilon_s)+{{p_1-p_2\over 2}}-4\Bigg)~,
\label{ND}
\ee
which generalizes the 9-dimensional result,  Eq.~(\ref{N9}). As a remark, we see that for $p_1=2, 3$, which correspond to $SO(2)$ and $[SO(3)\times SO(1)]$ gauge factors, the potential turns out to be independent of the single open string WL (up to exponentially suppressed terms), as is the case in region~(\ref{hyp}). 

Notice that Eq.~(\ref{ND}) is valid at arbitrary point in the open string moduli space  \ie that the $\varepsilon_r^I$'s are not assumed to be small. To discuss the stability of the backgrounds where all branes are located at corners (except when $p_1=2,3$, for which they can sit anywhere), it is however enough to Taylor expand $\N_{\tilde l}(\vec \W)$, which leads to 
\begin{align}
\!\!\!\!\V~=~ {\Gamma(5)\over \pi^{D+5}}\, {M_s^D\over 2^D} \sqrt{\det G} \,  \bigg\{ \big(\nF-\nB\big) \Xi  ~+~  & 4\pi^2 (p_1-2-p_2)\sum_{r=1}^{\half(p_1-p_2)}\varepsilon_r^I\Delta_{IJ}\varepsilon_r^J~+~\O(\varepsilon^4)\bigg\} \nonumber \\
&  ~+~\O\big(M_s^D\sqrt{\det G}\, G_{99}^{-{11\over 4}}e^{-2\pi\sqrt{G_{99}}}\big)~,
\end{align}
where the massless spectrum counting reproduces Eq.~(\ref{nfb0}), and where 
\be
\nF\,-\,\nB~=~8\Big(8-\half(p_1-p_2)^2\Big)~,~~ \Xi~=~ \sum_{\vec l}{1\over (\tilde l_KG_{KL}\tilde l_L)^5}~, ~~ \Delta_{IJ}~=~\sum_{\vec l}{\tilde l_I\tilde l_J\over (\tilde l_KG_{KL}\tilde l_L)^5}~.
\ee
Of course, the mass terms are absent for $p_1=0,1,2,3$. For $p_1\ge 4$, the WLs have positive definite  $r$-independent squared masses, if $\Delta_{IJ}$, $I,J=D,\dots,9$ is itself positive definite. 
This is easily seen to be the case, since $ V_I\Delta_{IJ}  V_J$ for an arbitrary vector $V_I$, yields 
\be
\Delta_{IJ} V_I V_J ~=~\sum_{\vec l}{(\tilde l_IV_I)^2 \over (\tilde l_KG_{KL}\tilde l_L)^5}  ~\geq~ 0 ~ , \label{pos1}
\ee
where $\tilde l_K G_{KL}\tilde l_L > 0$ since the metric   $G_{KL}$
is positive definite.

Minimizing the potential by setting these terms to zero, we then have
\be
\V~=~{\Gamma(5)\over \pi^{D+5}}\, {M_s^D\over 2^D} \sqrt{\det G}  \big(\nF-\nB\big) \Xi ~+~\O\big(M_s^D\sqrt{\det G}\, G_{99}^{-{11\over 4}}e^{-2\pi\sqrt{G_{99}}}\big)~,
\ee
whose dominant term depends on $G_{IJ}$ through $\det G$ and $\Xi$. It is therefore a source for the metric, unless $\nF-\nB=0$, \ie $(p_1,p_2)=(4,0)$ or $(5,1)$. As shown in Table~\ref{models}, the associated open string  gauge groups are
\be
SO(4)\times\big[SO(1)\times SO(1)\big]^{14}     \quad  \mbox{and}\quad   
  \big[SO(5)\times SO(1)\big]\times\big[SO(1)\times SO(1)\big]^{13} , 
  \label{solmax}
\ee
which can be realized in dimension $D\le 5$. Hence, at 1-loop, the above backgrounds yield massive open string WLs and marginal NS-NS moduli $G_{IJ}$ (including the supersymmetry breaking scale $M$), with a potential that is independent of the RR moduli $C_{IJ}$.  

Actually, when  $\nF-\nB\neq 0$, we may focus on the regime where all $G_{ii}$ are greater than $G_{99}$, with moderate non-diagonal metric components, which yields
\be
\sqrt{\det G}\, \Xi~\sim~ {u_D\dots u_8\over G_{99}^{D\over 2}}\sum_{l_9} {1\over (2l_9+1)^{10}}~,\quad \where\quad u_i~=~\sqrt{G_{ii}\over G_{99}}~,\;\; i\,=\,D,\dots, 8.
\ee
In that case, it is true that the complex structures $u_i$ cannot be stabilized at large values. Of course, $\sqrt{\det G}\, \Xi$ becomes proportional to  $(G^{99})^{D\over 2}$ in the moduli space region defined in Eq.~(\ref{hyp}). Thus, one may ask whether $\sqrt{\det G}\,\Xi$ can be minimized in the intermediate regime $G_{ij}=\O(G_{99})$, $|G_{i9}|\lesssim \sqrt{G_{99}}$. A numerical study shows that this is not the case at least for $D=8$.  

Schematically, the 1-loop stability of the models presented in Sect.~\ref{vefD}  applies when the internal metric components are in the shell comprised between $G^{99}$ and $G_{99}$. By contrast, the results of the present  subsection are useful when some of $G_{ii}$, $i=D,\dots,8$, are of the order of $G_{99}$, or greater. However, one could also have considered the effective potential in a ``mixed form'', with Lagrangian formulation for the internal lattice in direction 9 and only {\it some} of the other directions $i$. In that case, we would have shown the marginal stability of more models in the regime where the associated $G_{ii}$ are of the order of $G_{99}$, or greater, and all other components of the internal metric are in the shell  comprised between $G^{99}$ and $G_{99}$. 

One may also consider extending the domain of (marginal) stability to regions where some of the $G_{ii}$ are lower than, or of the order of $G^{99}$. In such a regime, the light open strings (in the original type~I picture) have corresponding momenta $m_i=0$. For instance, if all $G_{ii}$ are lower than $G^{99}$, then the light open strings must be massless. On the contrary, the closed string sector contains infinite towers of winding modes arising from the small directions $i$. As a result, the dressing function of the closed string sector contribution to the potential depends on the small $G_{ii}$, while it does not for its  open string counterpart. Hence, there is no possible exact compensation of the $8\times 8$ winding towers and we do not expect small $G_{ii}$ to be stable. 

\section{Nonperturbative analysis of the models}
\label{sec.4}

There are consistency conditions on string backgrounds of a nonperturbative nature that are invisible in string perturbation theory.  
One of them is the fact that, whereas in perturbation theory the 
ten-dimensional gauge group of the type~I theory looks to be $O(32)$ rather than $SO(32)$, at a nonperturbative level the part disconnected from $SO(32)$ cannot be defined~\cite{Witten:1998cd}. This is consistent with the fact that the dual heterotic string has a gauge group that is $Spin(32)/\Z_2$, which contains
in particular spinorial representations under the gauge group. More generally, there are nonperturbative consistency conditions of K-theory origin ~\cite{Witten:1998cd,Minasian:1997mm}, which can also be understood with simpler methods in  terms of consistency of gauge theories on various D-brane probes  \cite{Uranga:2000xp} from the viewpoint of local and global  \cite{Witten:1982fp} anomaly cancelations. 

Let us discuss which Wilson lines are allowed from this nonperturbative point of view.  Starting from $SO(32)$, continuous Wilson lines can be understood as a field theory breaking  and do not
present any subtleties. Potential problems can arise when we are considering discrete deformations (\ie which cannot be realized via a standard Higgs/Hosotani mechanism) with D-branes which have
orthogonal gauge groups, which correspond in a T-dual picture to branes frozen at orientifold fixed points. These objects are clearly crucial
in our constructions in the previous sections, since they are needed for the construction of stable configurations at the quantum level. The original argument in  Ref.~\cite{Witten:1998cd} can be slightly adapted  to  our case. Indeed,
in \cite{Witten:1998cd}  $O(N) \subset O(32)$ instantonic configurations break the gauge group to $O(32-N)$ and consistency problems arose from the fermions in the representation $(N,32-N)$
of $O(N) \times O(32-N)$. In our case, due to the Scherk-Schwarz supersymmetry breaking, these fermions are massive. However,  in all our configurations, and in order to increase the scalar potential, 
there are massless fermions in bifundamental representations of gauge groups due to the combined action of the supersymmetry breaking and  the Wilson lines.  

In order to illustrate the point, we start from the simplest examples in nine and eight dimensions. 
In nine dimensions, such configurations are of the form $SO(p_1) \times SO(p_2)$, with $p_1+p_2=32$, which is fixed by the RR tadpole condition.  The corresponding brane configuration can be 
described by a $32 \times 32 $ Wilson line matrix $\W = {\rm diag} (I_{p_1},  -I_{p_2})$. Its determinant is $\det \W = (-1)^{p_2}$, which implies that for $p_2$ even  $\W$  belongs to $SO(32)$, whereas
for $p_2$ odd it belongs to $O(32)$, but not to $SO(32)$.  In particular, in addition to the  trivial  $SO(32)$ brane configuration, the second stable configuration discussed in Sect.~\ref{warmup}, with gauge group
$SO(31) \times SO(1)$,  is realized with a WL matrix in $O(32)$ and not in $SO(32)$.  

In eight dimensions, one can add two Wilson lines, along the two  cycles of the toroidal internal space.  It is simpler to visualize the relevant brane configurations after two T-dualities, turning D9-branes
into D7-branes sitting at the four O7-fixed points. The generic configuration of this type has a gauge group $[SO(p_1) \times SO(p_2)] \times [ SO(p_3)  \times SO(p_4)] $, with
$p_1+p_2+p_3+p_4=32$.  Writing the Wilson lines in terms of the $p_i \times p_i$ matrix blocks,  they read
 \be
 \W_9 = {\rm diag} (I_{p_1},  -I_{p_2} , I_{p_3},  -I_{p_4} ) \quad, \quad \W_8 = {\rm diag} (I_{p_1},  I_{p_2} , - I_{p_3},  -I_{p_4} ) \ . \label{np1} 
 \ee
 Their determinants are given by  $\det W_9 = (-1)^{p_2+p_4}$, $\det W_8 = (-1)^{p_3+p_4}$. As a consequence, among the non-trivial stable
  eight-dimensional brane configurations,
 $[SO(29) \times SO(1)]\times [SO(1) \times SO(1)]$ belongs to $SO(32)$, whereas $SO(30) \times [SO(1) \times SO(1)]$ belongs {to  $O(32)$  and cannot therefore be defined nonperturbatively.} 
 
 The natural question is then to ask which of the brane configurations/Wilson lines in Table 2 are nonperturbatively allowed from this point of view. By compactifying our type~I models to $D$ dimensions,  there
 are (in a  T-dual language, obtained by dualizing all $10-D$ internal  coordinates) a number of $2^{10-D}$  O$(D-1)$  orientifold planes, on which one can have  $p_A$ coincident $\half$-branes, $A=1, \dots,   2^{10-D}$. 
 There are $10-D$  Wilson line matrices $\W_I$, of determinant
 \be
 \det \W_I = (-1)^{\sum_A p_A^{(I)}} \ , \quad I=D,\dots,9\; , 
 \label{np2}
   \ee 
 where $ p_A^{(I)}$ are branes {localized} in the $9-D$ hyperplane perpendicular to the internal coordinate $\tilde X^I$ and which is not passing through the origin of the ``internal box'' (there is a second hyperplane perpendicular  to $\tilde X_I$, which passes through the origin).   
 The conditions to be satisfied in order to select Wilson line matrices in  $SO(32)$ is therefore
 \be
 \sum_A p_A^{(I)} \in 2\integer  \ , \quad I=D,\dots, 9\; .
 \label{np3}
   \ee   
There seems to be enough freedom in the models of Table~\ref{models} to satisfy these constraints  by suitably distributing the minus signs among the discrete WLs.
      
Finally, another potential constraint comes from adding D5-brane probes into our models, which have $USp(2n)$ gauge groups, and then checking potential global Witten anomalies~\cite{Witten:1982fp}. However since the corresponding
 spectra are non-chiral {after compactification to four dimensions}, we did not find any additional constraints.         
    
\section{Comments on swampland conjectures}
\label{sec.5}

One natural application of the class of models we constructed in this paper is to test the various recent swampland conjectures \cite{Vafa:2005ui,ArkaniHamed:2006dz, Ooguri:2006in,Ooguri:2016pdq,Obied:2018sgi,Agrawal:2018own,Garg:2018reu,Ooguri:2018wrx}. In this section we make preliminary remarks and leave a full study to future work. 

$\bullet$
One of the swampland conjectures is that $|\V'| > C \V$, where $C$ is a constant of order~1~\cite{Obied:2018sgi}.
For the models with potentials that are not exponentially suppressed, since the potential is of runaway type in the supersymmetry breaking radius, this is always satisfied.    
The models with exponentially small effective potential, $\V \sim e^{- R} $, where $R$ is the typical Scherk-Schwarz radius, are somewhat different. The canonically normalized field is of the form $\sigma = \log R$. Then $|\V'|/\V \sim |R \V_R| /\V \sim R $ which becomes arbitrarily large, easily satisfying any constraint for large enough $R$. At higher-loop orders one may need an additional condition at each loop to cancel the leading contribution to the vacuum energy, so presumably 
 at some loop order the potential will become polynomial in $R$ and therefore $C = \O(1)$. 
 
 $\bullet$ Another swampland conjecture is that  the only possibility for the dark energy in string theory is quintessence  \cite{Agrawal:2018own}.  However,  whereas one can (relatively) easily find stable string models with positive (exponential for canonically normalized fields) potentials and runaway rolling vacua,  they do not generically lead to accelerating cosmologies. The reason  is that the exponent of the exponential is larger than the critical value (equal to $\sqrt{2}$ in Planck units in four-dimensions) needed to generate an accelerating universe.  It would be interesting to check if in more sophisticated compactifications
 with supersymmetry breaking, the universe is accelerating.

  $\bullet$ It would be worth investigating whether the generic nonperturbative  instability of the non-supersymmetric Kaluza-Klein vacua  
\cite{Witten:1981gj} takes place in our models. The latter possessing massless fermions, it is unclear {\it a priori} if the instability  persists.

$\bullet$ Finally, it would be interesting to study the weak gravity conjecture coming from brane-brane interactions,  and the quantum corrections to the D1-branes tensions and charges 
 in our class of models with positive scalar potential, by generalizing the framework recently discussed in \cite{Bonnefoy:2018mqb}.


\section{Conclusions and perspectives}
\label{sec.6}

In this paper, we presented a large class of models with exponentially small or positive effective potential in type~I string theory, at the 1-loop level. The models are based on simple toroidal 
compactifications, with discrete deformations corresponding to $\half$-branes stuck on orientifold fixed points (in a T-dual language). The great advance over previous works is that these models are (marginally) stable at 1-loop with respect to all moduli  fields, except the supersymmetry breaking scale and dilaton when the potential is non-vanishing (up to exponentially suppressed terms). To be specific, the open string Wilson lines have positive squared masses or are marginally stable, while the closed string NS-NS and RR moduli are flat directions at 1-loop. 

The essential ingredient of stuck (or half) D-branes  at orientifold fixed points has two simultaneous effects. On the one hand, in the presence of supersymmetry breaking, it ensures the presence in the massless spectrum of fermions stretched between pairs of O-planes  separated along the direction generating the Scherk-Schwarz supersymmetry breaking. On the other hand, the $\half$-branes do not introduce continuous Wilson line moduli which, if they existed, would generate instabilities, due to brane-brane attractions
generated by supersymmetry breaking. Such Wilson lines are not continuous deformations of the $SO(32)$ type~I superstring, but are rather discrete deformations contained either in $SO(32)$, or in a 
disconnected  component  of $O(32)$.  The configurations descending from $SO(32)$ should have a heterotic dual according to the adiabatic argument of Ref.~\cite{Vafa:1995gm}. An interesting exercise which we leave for future work would be to construct these stable heterotic duals explicitly. 

As the class of models we constructed  relies heavily on $\half$-D-branes  at orientifold fixed points, with no associated
gauge group, the largest possible gauge symmetry we can obtain is rather small : for a stable brane configuration with zero or positive scalar potential it is $SO(5)$, which is obviously not large enough to accommodate the Standard Model
gauge group. It is therefore an important question to find ways to enhance the available gauge symmetry without re-introducing Wilson line instabilities. One obvious way to do this would be to compactify on orbifolds. In this case, additional orientifold planes (O5-planes in type~I string,
which are of three different types) would be generated and corresponding D5-branes would have to be added, for consistency with the RR tadpole cancellation conditions.  In such a construction, the Standard Model
gauge group would then be realised on the D5-branes, with the D9-sector we have been focussing on in the present paper playing the role of a hidden sector generating the observed dark energy.

Finally, the class of open string models we have considered  extends that found in a heterotic context, and can be considered from a cosmological viewpoint. It turns out that whatever $\nF-\nB$ is, a flat, homogeneous and isotropic universe can always enter into an ever-expanding ``Quantum No-Scale Regime''~\cite{CFP,CP,Partouche:2018ftj}. What is meant by this is that the evolution approaches that found for $\nF-\nB=0$, thus restoring dynamically the no-scale structure \ie the flatness of the modulus $M$. Hence, once entering into such a regime, the characteristics of the potential are irrelevant, the latter being dominated by moduli kinetic energy. The sign of $\nF-\nB$ is however crucial in the sense that when it is positive, the evolutions are globally attracted towards such a  Quantum No-Scale Regime, while  if $\nF-\nB< 0$ this is only true at the price of imposing a relatively severe fine tuning of the initial conditions, in order to avoid a collapsing evolution. The moduli stability analyzed in our work may be relevant in this cosmological context once the models are rich enough to put a halt to the time-evolution of the supersymmetry breaking scale $M$, which we hope to address in future work. 


 \section*{Acknowledgements}
 
We are grateful to  Carlo Angelantonj for fruitful discussions. 
This work is supported by Royal-Society/CNRS International Cost Share Award IE160590. 
E.D. acknowledges partial financial support from the ANR Black-dS-String. D.L. is supported by an STFC studentship.


\section*{Appendix: 1-loop effective potential}
\label{A0}
\renewcommand{\theequation}{A.\arabic{equation}}
\renewcommand{\thesection}{A}
\setcounter{equation}{0}

The goal of this Appendix is to present in some detail the computation of the effective potential in the open string models considered in the core of the paper, at weak string coupling{\footnote{ For original
constructions see \cite{orientifolds,bps}. For  reviews, see \eg  \cite{reviews}}. 
In arbitrary dimension $D$, its expression may be divided into the contributions arising from the torus, Klein bottle, annulus and M\"obius strip amplitudes, 
\be
\begin{aligned}
&\V~=~-{M_s^D\over 2(2\pi)^D}\, (\T+\K+\A+\M)~,\espD \\
\where\quad & \T~=~\int_\F{d\tau_1d\tau_2\over \tau_2^{1+{D\over 2}}}\, \Str  q^{L_0-\half}\bar q^{\tilde L_0-\half}~,  &&\!\!\!\!\!\K=\int_0^{+\infty}{d\tau_2\over \tau_2^{1+{D\over 2}}}\, \Str  \Omega q^{L_0-\half}\bar q^{\tilde L_0-\half}~,\\
&\A~=~\int_0^{+\infty}{d\tau_2\over \tau_2^{1+{D\over 2}}}\, \Str  q^{\half(L_0-\half)},&& \!\!\!\!\!\!\!\M=\int_0^{+\infty}{d\tau_2\over \tau_2^{1+{D\over 2}}}\, \Str  \Omega q^{\half(L_0-\half)}~.
\end{aligned}
\label{Vdef}
\ee
In the above formula, $\tau_1, \tau_2$ are the real and imaginary parts of the Teichm\"uller parameter $\tau$, $q=e^{2i\pi\tau}$, $\F$ is the fundamental domain of $SL(2,\Z)$, $L_0,\tilde L_0$ are the zero frequency Virasoro operators, and $\Omega$ is the orientifold generator.  


\subsection{Conventions and notations}
\label{A1}

In type~I string theory compactified on a torus $T^{10-D}$, the amplitudes can be expressed in terms of lattices of zero modes and characters for the oscillators. Our notations are as follows~: 


\paragraph{\em Lattices~: } 

For the genus-1 Riemann surface, the expression of $\T$ involves 
\be
\begin{aligned}
&\Lambda_{\vec m, \vec n}(\tau)~=~q^{{1\over 4}P^L_IG^{IJ}P^L_J}\, \bar q^{{1\over 4}P^R_IG^{IJ}P^R_J}~, \\
&P^L_I~=~m_I+G_{IJ}n_J ~,\quad P^R_I\,\,=\,\,m_I-G_{IJ}n_J~, \quad  I=D,\dots, 9 ~,\esps
\end{aligned}
\ee
where $m_I, n_I$ are the momentum and winding numbers along the compact direction $X^I$, $G_{IJ}$ is the internal torus metric and $G^{IJ}=G^{-1}_{IJ}$. Due to the orientifold projection, the antisymmetric tensor $B_{IJ}$ present in the type~IIB string is absent.  

The closed strings running in the Klein bottle, as well as the type~I open strings  in the annulus and M\"obius strip have no winding modes for the background with D9 branes / O9 planes. Hence, it is natural to define the lattice of KK modes 
\begin{equation}
P_{\vec m}(i\tau_2)~=~\Lambda_{\vec m,\vec 0}(\tau)~=~e^{-\pi\tau_2m_IG^{IJ}m_J}~.\label{eq:P9dDef}
\end{equation}
When $\K$, $\A$, $\M$ are written in the closed string tree-level channel, they involve winding sums of the form 
\be
W_{\vec n}(i\ell)~=~\Lambda_{\vec 0,\vec n}(i\ell)~=~e^{-{\pi\over 2} \ell n_IG_{IJ}n_J}~.
\label{eq:WdDimDef}
\ee
One passes from one picture to the other by Poisson resumming, 
\begin{equation}
\!\!\!\sum_{\vec m}P_{\vec{m}+\vec{a}}(i\tau_2)~=\begin{cases}
(2\ell)^{10-D\over 2}\sqrt{\det G}\, \dis \sum_{\vec n} e^{2\pi i\vec{n}\cdot\vec{a}}\, W_{2\vec{n}} (i\ell)~, \mbox{where \;$\ell\,=\,\dis \frac{1}{2\tau_{2}}$\; for $\K$, $\M$,}\\
\Big(\dis {\ell\over 2}\Big)^{10-D\over 2}\sqrt{\det G}\, \sum_{\vec n} e^{2\pi i\vec{n}\cdot\vec{a}}\, W_{\vec{n}}(il)~, \;\mbox{where \;$\ell\,=\,\dis \frac{2}{\tau_{2}}$\;  for $\A$~.}\esp
\end{cases}\label{eq:PTransform}
\end{equation}


\paragraph{\em Characters~: } Our definitions of the Jacobi modular forms and Dedekind function are
\be
\label{th}
\theta\big[{}^\alpha_\beta\big](v| \tau)~=~\sum_m q^{{1\over 2}(m-{\alpha\over 2})^2}e^{2i\pi(v-{\beta\over 2})(m-{\alpha\over 2})} ~, \;\;\quad \eta(\tau)~=~ q^{1\over 24}\prod_{n=1}^{+\infty}(1-q^n)~.
\ee
At $v= 0$, it is standard to denote $\theta\big[{}^0_0\big]=\theta_3$, $\theta\big[{}^0_1\big]=\theta_4$, $\theta\big[{}^1_0\big]=\theta_2$, $\theta\big[{}^1_1\big]=\theta_1$, in terms of which the $SO(8)$ affine characters can be written as 
\begin{equation}
O_{8}~=~\frac{\theta_{3}^{4}+\theta_{4}^{4}}{2\eta^{4}}~,\quad \;\;V_{8}~=~\frac{\theta_{3}^{4}-\theta_{4}^{4}}{2\eta^{4}}~,\quad\;\; S_{8}~=~\frac{\theta_{2}^{4}+\theta_{1}^{4}}{2\eta^{4}}~,\quad\;\; C_{8}~=~\frac{\theta_{2}^{4}-\theta_{1}^{4}}{2\eta^{4}}~.\label{eq:CharacterDef}
\end{equation}
For the amplitudes $\T$, $\K$ and $\A$, the useful modular transformations are 
\begin{equation}
\begin{pmatrix}O_{8}\\
V_{8}\\
S_{8}\\
C_{8}
\end{pmatrix}\!\!(\tau)~=~\frac{1}{2}\begin{pmatrix}1 & 1 & 1 & 1\\
1 & 1 & -1 & -1\\
1 & -1 & 1 & -1\\
1 & -1 & -1 & 1
\end{pmatrix}\begin{pmatrix}O_{8}\\
V_{8}\\
S_{8}\\
C_{8}
\end{pmatrix}\!\!\Big(\!\!-\frac{1}{\tau}\Big)~, \qquad \eta(\tau)~=~{1\over \sqrt{-i \tau}}\, \eta\Big(\!\!-\frac{1}{\tau}\Big)~.
\label{eq:CharacterTransform}
\end{equation}

For the M\"obius strip amplitude, it is convenient to switch from any character $\chi$ to a real ``hatted'' character $\hat \chi$ defined by \cite{reviews} 
\begin{equation}
\hat{\chi}\Big(\frac{1}{2}+i\tau_{2}\Big)~=~e^{-i\pi(h-{c\over 24})}\, \chi\Big(\frac{1}{2}+i\tau_{2}\Big)~,
\end{equation}
where $h$ is the weight of the associated primary state and
$c$ is the central charge. The transformation from the open to the closed string channel, called the P-transformation, then takes the form
\begin{equation}
\begin{pmatrix}\hat O_{8}\\
\hat V_{8}\\
\hat S_{8}\\
\hat C_{8}
\end{pmatrix}\!\!\Big(\frac{1}{2}+i{\tau_2\over 2}\Big)=\diag(-1,1,1,1)\!\begin{pmatrix}\hat O_{8}\\
\hat V_{8}\\
\hat  S_{8}\\
C_{8}
\end{pmatrix}\!\!\Big(\frac{1}{2}+i\ell\Big)~, \quad \hat \eta\Big(\frac{1}{2}+i{\tau_2\over 2}\Big)=\sqrt{2\ell}\, \hat \eta\Big(\frac{1}{2}+i\ell\Big)~.
\label{eq:hat chi}
\end{equation}


\paragraph{\em Limiting behaviours~: } In the final expressions of the amplitudes, we display the dominant contributions arising from light states (compared to the supersymmetry breaking scale) and are more schematic about those associated with heavy modes. For this purpose, we will use 
\be 
\label{dexpsup}
\Hc_\nu(z)~=~ {1\over \Gamma(\nu)}\int_0^{+\infty}{dx\over x^{1+\nu}}\, e^{-{1\over x}-z^2x}~=~{2\over \Gamma(\nu)}\, z^\nu K_\nu(2z)~,
\ee
where $K_\nu$ is a modified Bessel function of the second kind. At large and small arguments, it has the following behaviour :   
\be 
\label{expsup}
\Hc_\nu(z)~\sim~ {\sqrt{\pi}\over \Gamma(\nu)}\, z^{\nu-{1\over 2}}\, e^{-2z} \mbox{~as} \;\;\; z\gg 1~,~~~ \Hc_\nu(z)~=~1-{z^2\over \nu-1}+\O(z^4)\;\;\; \mbox{as}\;\;\; |z|\ll 1~.
\ee


\subsection{Massless spectrum and potential in 9 dimensions}
\label{A2}

We are interested in the orientifold projection of the type~IIB theory in 9 dimensions, with Scherk-Schwarz spontaneous breaking of supersymmetry implemented along the internal circle $S^1(R_9)$ of radius $R_9$. The torus amplitude contribution to the effective potential $\V$ is
\be
\begin{aligned}
\!\T~=~ {1\over 2}\int_\F{d^2\tau\over \tau_2^{11\over 2}}\,{1\over \eta^8\bar \eta^8} &\sum_{m_9,n_9}\!\Big\{\big(V_8\bar V_8+S_8\bar S_8\big)\Lambda_{m_9,2n_9}- \big( V_8\bar S_8+S_8\bar V_8\big)\Lambda_{m_9+\half,2n_9}  \\
&~+~\big(O_8\bar O_8+C_8\bar C_8\big)\Lambda_{m_9,2n_9+1}- \big( O_8\bar C_8+C_8\bar O_8\big)\Lambda_{m_9+\half,2n_9+1}\Big\}~,
\end{aligned}
\ee
where the lattices depend on $G_{99}=R_9^2$. The orientifold projection leads to the overall  normalization factor $\half$, as well as to the Klein bottle contribution 
\begin{equation}
\mathcal{K}~=~\half\int_0^{+\infty}{d\tau_2\over \tau_2^{11\over 2}}\, {1\over \eta^8} \sum_{m_9}\,(V_{8}-S_{8}) P_{m_9}~,
\end{equation}
where the argument of the characters is $2i\tau_2$. As explained in Sect.~\ref{GS}, the open string sector can be described either in type~I or type~I' language, obtained by T-dualizing $S^1(R_9)$. In type~I', the $\alpha$-th D8-brane is located at $2\pi a_\alpha \tilde R_9$ along the dual circle, where $\tilde R_9=1/R_9$. 


\paragraph{\em Spectrum~: } 

For reasons that will become clear shortly, in order to determine the massless spectrum, we first split the generic configuration as follows~:

$\bullet$ $p_1$ $\half$-branes on an O8-orientifold plane located at $a=0$,

$\bullet$ $p_2$ $\half$-branes on a second O8-orientifold plane located at $a=\half$,

$\bullet$ $q$ branes at $a={1\over 4}$, with their mirrors at $a=-{1\over 4}$,

$\bullet$ $r_\sigma$ branes at $a=a_\sigma\in\; (0,{1\over 4})\; \cup\;  ({1\over 4},\half)$, with their mirrors at $a=-a_\sigma$.  

\noindent Notice that $p_1$ and $p_2$ can be even or odd. By denoting the degeneracies $q,r_\sigma$ by $\bar q,\bar r_\sigma$ when the momentum shifts of $m_9$ are $-{1\over 4}$, $-a_\sigma$, the annulus amplitude reads\footnote{Open strings Wilson lines were originally introduced in \cite{bps}. In the context of Scherk-Schwarz models with open strings, this was done in \cite{open2_ss}.}  
\begin{align}
\mathcal{A} ~ =~&\, \frac{1}{2}\int_0^{+\infty}\frac{d\tau_{2}}{\tau_{2}^{11\over 2}}\, \frac{1}{\eta^{8}}\sum_{m_9} \Big\{\big(p_{1}^{2}+p_{2}^{2}+2q\bar{q}+2\sum_\sigma r_{\sigma}\bar{r}_{\sigma}\big)\big(V_{8}P_{m_9}-S_{8}P_{m_9+\frac{1}{2}}\big) \nonumber \\
 & +2p_{1}p_{2}\big(V_{8}P_{m_9+\frac{1}{2}}-S_{8}P_{m_9}\big)+q^{2}\big(V_{8}P_{m_9+\frac{1}{2}}-S_{8}P_{m_9}\big)+\bar{q}^{2}(V_{8}P_{m_9-\frac{1}{2}}-S_{8}P_{m_9}\big)\phantom{\sum_\sigma}\nonumber \\
 & +\sum_\sigma r_{\sigma}^{2}\big(V_{8}P_{m_9+2a_{\sigma}}-S_{8}P_{m_9+\frac{1}{2}+2a_{\sigma}}\big)+\sum_\sigma \bar{r}_{\sigma}^{2}\big(V_{8}P_{m_9-2a_{\sigma}}-S_{8}P_{m_9+\frac{1}{2}-2a_{\sigma}}\big)\nonumber \\
 & +2p_{1}q\big(V_{8}P_{m_9+\frac{1}{4}}-S_{8}P_{m_9-\frac{1}{4}}\big)+2p_{1}\bar{q}\big(V_{8}P_{m_9-\frac{1}{4}}-S_{8}P_{m_9+\frac{1}{4}}\big)\phantom{\sum_\sigma}\nonumber \\
 & +2p_{2}q\big(V_{8}P_{m_9-\frac{1}{4}}-S_{8}P_{m_9+\frac{1}{4}}\big)+2p_{2}\bar{q}\big(V_{8}P_{m_9+\frac{1}{4}}-S_{8}P_{m_9-\frac{1}{4}}\big)\phantom{\sum_\sigma}\nonumber \\
 & +2\sum_\sigma p_{1}r_{\sigma}\big(V_{8}P_{m_9+a_{\sigma}}-S_{8}P_{m_9+\frac{1}{2}+a_{\sigma}}\big)+2\sum_\sigma p_{1}\bar{r}_{\sigma}\big(V_{8}P_{m_9-a_{\sigma}}-S_{8}P_{m_9+\frac{1}{2}-a_{\sigma}}\big)\phantom{\sum_\sigma}\nonumber \\
 & +2\sum_\sigma p_{2}r_{\sigma}\big(V_{8}P_{m_9+\frac{1}{2}+a_{\sigma}}-S_{8}P_{m_9+a_{\sigma}}\big)+2\sum_\sigma p_{2}\bar{r}_{\sigma}\big(V_{8}P_{m_9+\frac{1}{2}-a_{\sigma}}-S_{8}P_{m_9-a_{\sigma}}\big)\nonumber \\
 & +2\sum_\sigma qr_{\sigma}\big(V_{8}P_{m_9+\frac{1}{4}+a_{\sigma}}-S_{8}P_{m_9-\frac{1}{4}+a_{\sigma}}\big)+2\sum_\sigma q\bar{r}_{\sigma}\big(V_{8}P_{m_9+\frac{1}{4}-a_{\sigma}}-S_{8}P_{m_9-\frac{1}{4}-a_{\sigma}}\big)\nonumber \\
 & +2\sum_\sigma \bar{q}r_{\sigma}\big(V_{8}P_{m_9-\frac{1}{4}+a_{\sigma}}-S_{8}P_{m_9+\frac{1}{4}+a_{\sigma}}\big)+2\sum_\sigma\bar{q}\bar{r}_{\sigma}\big(V_{8}P_{m_9-\frac{1}{4}-a_{\sigma}}-S_{8}P_{m_9+\frac{1}{4}-a_{\sigma}}\big)\nonumber \\
 & +\sum_{\sigma\neq\tau} r_{\sigma}r_{\tau}\big(V_{8}P_{m_9+a_{\sigma}+a_{\tau}}-S_{8}P_{m_9+\frac{1}{2}+a_{\sigma}+a_{\tau}}\big)+\sum_{\sigma\neq\tau}\bar{r}_{\sigma}\bar{r}_{\tau}\big(V_{8}P_{m_9-a_{\sigma}-a_{\tau}}-S_{8}P_{m_9+\frac{1}{2}-a_{\sigma}-a_{\tau}}\big)\nonumber \\
 & +2\sum_{\sigma\neq \tau} r_{\sigma}\bar{r}_{\tau}\big(V_{8}P_{m_9+a_{\sigma}-a_{\tau}}-S_{8}P_{m_9+\frac{1}{2}+a_{\sigma}-a_{\tau}}\big)\Big\}~, 
\end{align}
where the argument of the characters is ${i\over 2}\tau_2$. Anticipating the RR tadpole cancellation condition, the M\"obius strip contribution is dressed with an overall minus sign, 
\begin{eqnarray}
\mathcal{M}  &~=~&-\frac{1}{2}\int_0^{+\infty}\frac{d\tau_{2}}{\tau_{2}^{11/2}}\, \frac{1}{\hat \eta^{8}}\sum_{m_9} \Big\{\big(p_{1}+p_{2})\big(\hat V_{8}P_{m_9}-\hat S_{8}P_{m_9+\frac{1}{2}}\big) \nonumber  \\
&&~+~~q\big(\hat V_{8}P_{m_9+\frac{1}{2}}-\hat S_{8}P_{m_9}\big)+\bar{q}\big(\hat V_{8}P_{m_9-\frac{1}{2}}-\hat S_{8}P_{m_9}\big) \phantom{\sum_\sigma} \\ 
 && ~+~\sum_\sigma r_{\sigma}\big(\hat V_{8}P_{m_9+2a_{\sigma}}-\hat S_{8}P_{m_9+\frac{1}{2}+2a_{\sigma}}\big)+\sum_\sigma\bar{r}_{\sigma}\big(\hat V_{8}P_{m_9-2a_{\sigma}}-\hat S_{8}P_{m_9+\frac{1}{2}-2a_{\sigma}}\big)\Big\}~,\nonumber 
\end{eqnarray}
with hatted characters at $\half+{i\over 2}\tau_2$. 

In the closed string sector, due to the Scherk-Schwarz mechanism, the massless states are bosons, which are present in the parent type~IIB theory in 10 dimensions. Those surviving the orientifold projection are the dilaton $\phi$, metric $G_{MN}$ and RR 2-form $C_{MN}$, which yield $1+35+28=8\times 8$ states. In the open string sector, the massless bosons and fermions contributing to $\A+\M$ are respectively enumerated by the degeneracies of the $(V_8/\eta^8)P_0$ and $-(S_8/\eta^8)P_{0}$ blocks (or their hatted counterparts). Expanding $V_8/\eta^8=S_8/\eta^8=8(1+\O(q))$ (and similarly for the hatted characters), the numbers of massless bosons and fermions are  given by
\be
\begin{aligned}
n_{B}^{(0)} ~=~&\;8\bigg(8~+~\frac{p_{1}(p_{1}-1)}{2}+\frac{p_{2}(p_{2}-1)}{2}~+~q\bar{q}~+~\sum_\sigma r_{\sigma}\bar{r}_{\sigma}\bigg), \\
n_{F}^{(0)} ~=~& \;8\bigg(p_{1}p_{2}~+~\frac{q(q-1)+\bar{q}(\bar{q}-1)}{2}~+\sum_{\overset{\scriptstyle \sigma<\tau}{a_{\sigma}+a_{\tau}=\half}}(r_{\sigma}r_{\tau}+\bar{r}_{\sigma}\bar{r}_{\tau})\bigg)~.
\label{nfbtot}
\end{aligned}
\ee
As a result, the open string massless states amount to the bosonic parts of vector multiplets in the adjoint representation of  $SO(p_{1})\times SO(p_{2})\times U(q)\times\prod_{\sigma}U(r_{\sigma})$, and fermionic parts of vector multiplets  in the bifundamental of $SO(p_1)\times SO(p_2)$, in the antisymmetric $\oplus$ $ \overline{\mbox{antisymmetric}}$ of $U(q)$, and in the bifundamental $\oplus$ $\overline{\mbox{bifundamental}}$ of $U(r_\sigma)\times U(r_\tau)$, when accidentally $a_\sigma+a_\tau=\half$. 


\paragraph{\em Effective potential~: } 

We proceed with the derivation of the 1-loop effective potential. For this purpose, it is convenient to define a WL matrix 
\be
\W~=~\diag\!\Big(e^{2i\pi a_\alpha}; ~\alpha\,=\,1,\dots,p_1+p_2+2q+2\sum_\sigma r_\sigma  \Big)~,
\ee
where $a_\alpha$ is the position of the $\alpha$-th $\half$-brane, and to write the open string channel amplitudes as 
\be
\begin{aligned}
\mathcal{A} & ~=~\frac{1}{2}\int_{0}^{\infty}\frac{d\tau_{2}}{\tau_{2}^{11\over 2}}\,~ \frac{1}{\eta^{8}} \sum_{m_9}\sum_{\alpha,\beta} \big(V_{8}P_{m_9+ a_{\alpha}-a_{\beta}}-S_{8}P_{m_9+\frac{1}{2}+a_{\alpha}-a_{\beta}}\big)~, \\
\mathcal{M} & ~=~-\frac{1}{2}\int_{0}^{\infty}\frac{d\tau_{2}}{\tau_{2}^{11\over 2}}\, ~\frac{1}{\hat{\eta}^{8}} \sum_{m_9}\sum_{\alpha}\big(\hat{V}_{8}P_{m_9+2 a_{\alpha}}-\hat{S}_{8}P_{m_9+\frac{1}{2}+2a_{\alpha}}\big)~.
\end{aligned}
\ee
These expressions can be written in the dual closed string channel. Using the Poisson resummation formulas~(\ref{eq:PTransform}) and  transformations~(\ref{eq:CharacterTransform}), (\ref{eq:hat chi}), we obtain
\begin{align}
\mathcal{A} & ~=~\frac{2^{-5}}{2}R_{9}\int_{0}^{\infty}\frac{d\ell}{\eta^{8}}\sum_{n_9}\left(\left(\text{tr}\, \mathcal{W}^{2n_9}\right)^{2}(V_{8}-S_{8})W_{2n_9}+\left(\text{tr}\,\mathcal{W}^{2n_9+1}\right)^{2}(O_{8}-C_{8})W_{2n_9+1}\right)~,\nonumber  \\
\mathcal{M} & ~=~-R_{9}\int_{0}^{\infty}{d\ell\over \hat \eta^8}\sum_{n_9}\big(\text{tr}\,\mathcal{W}^{2n_9}\big)\big(\hat{V}_{8}-(-1)^{n_9}\hat{S}_{8}\big)W_{2n_9}~,
\end{align}
where $\ell=\frac{2}{\tau_{2}}$ and $\ell=\frac{1}{2\tau_{2}}$
for the annulus  and M\"obius strip amplitudes, respectively. The arguments of the characters in $\A$ and $\M$ are $i\ell$ and $\half+i\ell$, respectively. Even though $\K$ vanishes, we may also write it in the transverse channel,  
 \begin{equation}
\mathcal{K}~=~{2^5\over 2}R_9\int_0^{+\infty}{d\ell\over \eta^8}\sum_{n_9} \,(V_{8}-S_{8}) W_{2n_9}~,
\end{equation}
where $\ell={1\over 2\tau_2}$ and the characters are taken at $i\ell$. There are no UV divergence as $\ell\to +\infty$ ($\tau_2\to 0$) in $\K+\A+\M$, when the RR tadpole cancellation condition is obeyed. The latter amounts to setting the coefficient of $(S_{8}/\eta^8)W_0$ (or $(\hat S_8/\hat \eta^8)W_0$) to zero. Besides the sign of $\M$ already mentioned, this constrains the number of $\half$-branes to be 
\begin{equation}
p_{1}+p_{2}+2q+2\sum_{\sigma}r_{\sigma}~=~32~.
\label{RRtad}
\end{equation}

In the large $R_9$ limit, the torus amplitude $\T$ is dominated by the level-matched pure KK modes associated to $S^1(R_9)$. The contributions of all oscillator states, winding modes, and the non-level matched states  are exponentially suppressed (this is shown in arbitrary dimension $D$ in Appendix~\ref{A3}). Using the behaviour of the function $\Hc_\nu$ in Eqs~\ref{expsup}), (\ref{dexpsup}), one obtains 
\begin{equation}
\mathcal{T}~=~\frac{\Gamma(5)}{\pi^{5}}\, {8\over R^9_{9}}\sum_{n_9}\frac{16}{(2n_9+1)^{10}}~+~\mathcal{O}\bigg({e^{-4\pi R_{9}}\over R_{9}^{9/2}}\bigg)~,
\end{equation}
where $n_9$ denotes for notational convenience the Poisson ressummed index of the momentum $m_9$.  In the same limit, $\A+\M$ in the closed string channel reads
\begin{equation}
\mathcal{A}+\mathcal{M}~=~\frac{\Gamma(5)}{\pi^{5}}\, {8\over R^9_{9}}\sum_{n_9}\frac{\left(\text{tr}\, \mathcal{W}^{2n_9+1}\right)^{2}-\text{tr}\, \mathcal{W}^{2(2n_9+1)}}{(2n_9+1)^{10}}~+~\mathcal{O}\bigg({e^{-2\pi R_{9}}\over R_{9}^{9/2}}\bigg)~.
\end{equation}
Hence, the total effective potential~(\ref{Vdef}) is  
\begin{equation}
\V~=~\frac{\Gamma(5)}{\pi^{14}}\, {M_s^9 \over (2R_9)^9}\, 4\sum_{n_9}\frac{-16-\left(\text{tr }\mathcal{W}^{2n_9+1}\right)^{2}+\text{tr }\mathcal{W}^{2(2n_9+1)}}{(2n_9+1)^{10}}~+~\mathcal{O}\bigg({M_s^9\over R_{9}^{9/2}}\, e^{-2\pi R_{9}}\bigg)~.
\label{eq:AppendixPotential}
\end{equation}


\subsection{Massless spectrum and potential in \bm $D$ dimensions}
\label{A3}

In this subsection, we extend some of the 9-dimensional results to the case of a toroidal compactification on $T^{10-D}$. The metric of the internal torus is $G_{IJ}$, and the Scherk-Schwarz mechanism is implemented along the direction $X^9$. The genus-1 Riemann surface amplitude is then 
\begin{align}
\T~&=~ {1\over 2}\int_\F{d^2\tau\over \tau_2^{D+2\over 2}}\,{1\over \eta^8\bar \eta^8} \sum_{\vec m,\vec n}\Big\{\big(V_8\bar V_8+S_8\bar S_8\big)\Lambda_{\vec m,(\vec n',2n_9)}- \big( V_8\bar S_8+S_8\bar V_8\big)\Lambda_{\vec m+\vec a_S,(\vec n',2n_9)}  \nonumber \\
&\qquad\qquad~+~\big(O_8\bar O_8+C_8\bar C_8\big)\Lambda_{\vec m,(\vec n',2n_9+1)}- \big( O_8\bar C_8+C_8\bar O_8\big)\Lambda_{\vec m+\vec a_S,(\vec n',2n_9+1)}\Big\}~,
\label{TD}
\end{align}
where $\vec a_S$ is the $(10-D)$-dimensional vector that implements the  $\half$-shift of the momentum $m_9$, while any ``primed''  vector only has $9-D$ entries corresponding to the non-Scherk-Schwarz directions,
\be
\vec a_S~=~\Big(\vec 0',\half\Big)~, \qquad \vec n~=~(\vec n',n_9)~.
\ee
The Klein bottle contribution is
\begin{equation}
\mathcal{K}~=~\frac{1}{2}\int_0^{+\infty}\frac{d\tau_{2}}{\tau_{2}^{D+2\over 2}}\,{1\over \eta^8}\sum_{\vec m} \, (V_{8}-S_{8})P_{\vec m}~.
\label{KD}
\end{equation}


\paragraph{\em Specific brane configuration and spectrum~: }

It is convenient to define the open string sector in the geometric type~II orientifold picture obtained by T-dualizing all internal directions. The initial D9-branes and O9-plane then translate into D$(D-1)$-branes with O$(D-1)$-planes. If the dual torus has metric $\tilde G_{IJ}=G^{IJ}$, we choose  to write the amplitudes in terms of the initial type~I metric $G_{IJ}$, in order to match with the closed string sector notations. 

In type~II orientifolds, there is one orientifold plane located at each corner of a $(10-D)$-dimensional box. The configurations  we are interested in consist of $p_A$ $\half$-branes located on the $A$-th O-plane, $A=1,\dots,2^{10-D}$. Their coordinates along the dual torus directions $\tilde X^I$ are $2\pi a_A^I\sqrt{\tilde G_{II}}$ (no sum over $I=D,\dots,9$), as shown in Fig.~\ref{dDim}. These positions can be encoded by WL vectors $\vec a_A$, whose components $a_A^I$, $I=D,\dots,9$, take discrete values 0 or $\half$. By convention, we choose  an ordering of the orientifolds planes such that  $\vec a_{2A}=\vec a_{2A-1}+\vec a_S$, $A=1,\dots, 2^{10-D}/2$. (Alternatively, we may write $\vec a_{2A-1}'=\vec a_{2A}'$.) In these notations, the open sector amplitudes can be written as 
\be
\begin{aligned}
\mathcal{A}&~=~\frac{1}{2}\int_0^{+\infty}\frac{d\tau_{2}}{\tau_{2}^{D+2\over 2}}\, {1\over \eta^8}\sum_{\vec m}\sum_{A,B=1}^{2^{10-D}}p_{A}p_{B}(V_{8}P_{\vec{m}+\vec{a}_{A}-\vec{a}_{B}}-S_{8}P_{\vec{m}+\vec{a}_{S}+\vec{a}_{A}-\vec{a}_{B}})~,\\  
\mathcal{M}&~=~ - \frac{1}{2}\int_0^{+\infty}\frac{d\tau_{2}}{\tau_{2}^{D+2\over 2}}\, {1\over \hat \eta^8}\sum_{\vec m}\sum_{A=1}^{2^{10-D}}p_{A}(\hat{V}_{8}P_{\vec{m}}-\hat{S}_{8}P_{\vec{m}+\vec{a}_{S}})~,
\end{aligned}
\ee
where the momenta shifts in $\M$ are trivial, $2a_A^I =0,1$. The number of open string massless states can be read off from the coefficients of $V_8P_{\vec 0}$ and $-S_8P_{\vec 0}$ (or their hatted counterparts). Massless bosons require $A=B$, while fermions are massless if and only if $\vec a_S+\vec a_A-\vec a_B=\vec 0$ or $2\vec a_S$.  Taking into account the closed string sector, we have in total  
\be
\nB ~=~8\bigg(8+\sum_{A=1}^{2^{10-D}} \frac{p_{A}(p_{A}-1)}{2}\bigg)~,\;\;\quad \nF ~=~8\sum_{A=1}^{2^{10-D}/2}{p_{2A-1}p_{2A}+p_{2A}p_{2A-1}\over 2}~.
\label{nfbtotD}
\ee
The open strings states amount to  the bosonic parts of vector multiplets in the adjoint representation of $\prod_A^{2^{10-D}} SO(p_A)$, coupled to the fermionic parts of vector multiplets in the bifundamentals of $SO(p_{2A-1})\times SO(p_{2A})$, $A=1,\dots,2^{10-D}/2$. As a result, we obtain
\be
\begin{aligned}
\nF-\nB &~=~8\,\bigg(-8-\half\sum_{A=1}^{2^{10-D}/2}(p_{2A-1}-p_{2A})^{2}+\frac{1}{2}\sum_{A=1}^{2^{10-D}}p_{A}\bigg) ~,\\
&~=~8\,\bigg(8-\frac{1}{2}\sum_{A=1}^{2^{10-D}/2}(p_{2A-1}-p_{2A})^{2}\bigg)~.
\label{spe}
\end{aligned}
\ee
In the second line, we use the RR tadpole cancellation condition, which fixes the number of $\half$-branes to be $\sum_{A=1}^{2^{10-D}}p_A=32$. This can be derived as in 9 dimensions from the amplitudes in the 
tree-level gravitational channel.


\paragraph{\em Effective potential at low supersymmetry breaking scale~: } 

Let us move on the computation of the 1-loop effective potential. For the time being, assume  that  the internal metric  induces only mass scales greater than the supersymmetry breaking scale. To be specific, we assume that 
\be
G^{99}\ll | G_{ij}|\ll G_{99}~, \;\;\quad |G_{9j}|\ll \sqrt{G_{99}}~,\;\;\quad i,j=D,\dots,8~,
\label{plateau}
\ee  
where $G_{99}\gg 1$ is understood, in order to avoid tachyonic instabilities.

In the open string sector, the WL moduli can be organized in matrices as
\be
\W_I~=~ \diag\!\Big(e^{2i\pi a^I_\alpha}; \alpha=1,\dots,32\Big)~,\quad I=D,\dots,9~ , 
\label{WI}
\ee
where $\alpha$ labels the $\half$-branes. At a generic point in moduli space, we will denote by $\vec a_\alpha$ the vectors with real entries $a_\alpha^I$, $I=D,\dots,9$. Of course, not all of them are independent dynamical degrees of freedom, since dynamical branes can freely move only in pairs with their images, while the remaining ones are frozen at O$(9-D)$-planes.  

In this notation, the annulus amplitude can be written 
\be
\mathcal{A}  ~=~\frac{1}{2}\int_{0}^{\infty}\frac{d\tau_{2}}{\tau_{2}^{D+2\over 2}}\, \frac{1}{\eta^{8}} \sum_{\vec m}\sum_{\alpha,\beta} (V_{8}P_{\vec m+ \vec a_{\alpha}-\vec a_{\beta}}-S_{8}P_{\vec m+\vec a_S+\vec a_{\alpha}-\vec a_{\beta}})~.
\label{ampA}
\ee
Expanding $V_8/\eta^8=S_8/\eta^8=8\sum_{k\ge 0}c_ke^{-\pi  k \tau_2}$, where $c_0=1$,  and Poisson resumming over $m_9$, we obtain
\be
\begin{aligned}
\A~=~ &\;  (G^{99})^{D\over 2}\, {\Gamma\big({D+1\over 2}\big)\over \pi^{D+1\over 2}}\, 8\sum_{k\ge 0}c_k\sum_{\alpha,\beta}\sum_{\vec m'}\sum_{l_9}{1\over |2l_9+1|^{D+1}} \\
&\; \cos\!\Big[2\pi(2l_9+1)\Big(a_\alpha^9-a_\beta^9+{G^{9i}\over G^{99}}(m_i+a_\alpha^i-a_\beta^i)\Big)\Big]\,  \Hc_{D+1\over 2}\Big(\pi|2l_9+1|{\M_{\A}\over \sqrt{G^{99}}}\Big)~,
\label{AD}
\end{aligned}
\ee
where the function $\Hc_\nu$ is given in Eq.~(\ref{dexpsup}). In the above expression, we have introduced a mass scale $\M_{\A}$ (in string units) that characterizes a KK tower of modes 
propagating  along the large \SS direction $X^9$, 
\be
\M^2_{\A}~=~ (m_i+a_\alpha^i-a_\beta^i)\hat G^{ij}(m_j+a_\alpha^j-a_\beta^j)+k~.
\ee
This definition involves the effective inverse metric of the internal space transverse to the \SS direction, 
\be
\hat G^{ij}~=~ G^{ij}-{G^{i9}\over G^{99}}\, G^{99}\, {G^{9j}\over G^{99}} ~=~ G^{ij}+\O\Big({1\over G_{99}}\Big), \quad i,j=D,\dots, 8~.\esp
\label{hG}
\ee
When the WL configuration describes  stacks of $p_A$, $A=1,\dots 2^{10-D}$,  $\half$-branes  located in the neighborhoods of the corners of the ``internal box'', we can split the WLs  into background values and deviations,
\be
a_\alpha^I~=~\langle a_\alpha^I\rangle+\varepsilon_\alpha^I~, \quad \where \quad \langle a_\alpha^I\rangle\in\Big\{0, \half\Big\}~, \quad \alpha=1,\dots,32~, \quad I=D,\dots, 9~.
\ee
In that case, $\M_{\A}=\O(1)$ unless $k=0$ and $m_i+\langle a_\alpha^i\rangle-\langle a_\beta^i\rangle=0$, $i=D,\dots,8$. This second condition amounts to  having $\vec m'=\vec 0'$ and $(\alpha,\beta)$ in the set $L$, such that 

$\bullet$ $\alpha, \beta$ belong to a bunch of $p_A$ $\half$-branes, $A=1,\dots, 2^{10-D}$, 

$\bullet$ or $\alpha, \beta$ belong respectively to bunches of $p_{2A-1}$ and $p_{2A}$ $\half$-branes,  $A=1,\dots, 2^{10-D}/2$,

$\bullet$ or $\beta, \alpha$ belong respectively to bunches of $p_{2A-1}$ and $p_{2A}$ $\half$-branes,  $A=1,\dots, 2^{10-D}/2$.

\noindent Due to the exponential suppression of the function $\Hc_{D+1\over 2}$ at large argument, we obtain
{\begin{align}
\A~=~ & \big(\sqrt{G^{99}}\big)^{D} {\Gamma\big({D+1\over 2}\big)\over \pi^{D+1\over 2}}\, 8\sum_{(\alpha,\beta)\in L}(-1)^{2(\langle a_\alpha^9\rangle -\langle a_\beta^9\rangle)}\sum_{l_9}{\cos\!\Big[2\pi(2l_9+1)\big(\varepsilon_\alpha^9-\varepsilon_\beta^9+{G^{9i}\over G^{99}}(\varepsilon_\alpha^i-\varepsilon_\beta^i)\big)\Big]\over |2l_9+1|^{D+1}}\nonumber \\
&\;\times \Hc_{D+1\over 2}\bigg(\pi|2l_9+1|{\big[(\varepsilon_\alpha^i-\varepsilon_\beta^i)\hat G^{ij}(\varepsilon_\alpha^j-\varepsilon_\beta^j)\big]^\half\over \sqrt{G^{99}}}\bigg)\!~+~\O\bigg(\big(\sqrt{G^{99}}\big)^{D\over 2}\, e^{- {2\pi c\over \sqrt{G^{99}}}}\bigg)~,
\label{ADs}
\end{align}}
where $c>0$ is moduli-dependent but $\O(1)$. 

The M\"obius strip amplitude
\be
\M  ~=~-\frac{1}{2}\int_{0}^{\infty}\frac{d\tau_{2}}{\tau_{2}^{D+2\over 2}}\, \frac{1}{\hat \eta^{8}} \sum_{\vec m}\sum_{\alpha} (\hat V_{8}P_{\vec m+ 2\vec a_{\alpha}}-\hat S_{8}P_{\vec m+\vec a_S+2\vec a_{\alpha}})
\label{ampM}
\ee
can be treated in a similar way, and yields
\be
\begin{aligned}
\M~=~ & - (G^{99})^{D\over 2}\, {\Gamma\big({D+1\over 2}\big)\over \pi^{D+1\over 2}}\, 8\sum_{k\ge 0}(-1)^kc_k\sum_{\alpha}\sum_{\vec m'}\sum_{l_9}{1\over |2l_9+1|^{D+1}} \\
&~~ \cos\!\Big[2\pi(2l_9+1)\Big(2a_\alpha^9+{G^{9i}\over G^{99}}(m_i+2a_\alpha^i)\Big)\Big]\,  \Hc_{D+1\over 2}\Big(\pi|2l_9+1|{\M_{\M}\over \sqrt{G^{99}}}\Big)~,
\end{aligned}
\label{MD}
\ee
where the KK tower mass scale satisfies
\be
\M^2_{\M}~=~ (m_i+2a_\alpha^i)\hat G^{ij}(m_j+2a_\alpha^j)+k~.
\ee
In this case one must pick the states satisfying $m_i+2\langle a_\alpha^i\rangle=0$ for the contributions not to be exponentially suppressed. However this fixes $m_i$ uniquely  to be 0 or 1. Ultimately we find
{\be
\begin{aligned}
\M~=~ &- \big(\sqrt{G^{99}}\big)^{D} \, {\Gamma\big({D+1\over 2}\big)\over \pi^{D+1\over 2}}\, 8\sum_{\alpha}\sum_{l_9}{\cos\!\Big[4\pi(2l_9+1)\big(\varepsilon_\alpha^9+{G^{9i}\over G^{99}}\, \varepsilon_\alpha^i\big)\Big]\over |2l_9+1|^{D+1}} \\
&\;\;\;\;\;\;\;\;\;\;\;\;\;\;\;\;\;\times  \Hc_{D+1\over 2}\bigg(2\pi|2l_9+1|{\big[\varepsilon_\alpha^i\,\hat G^{ij}\,\varepsilon_\alpha^j\big]^\half\over \sqrt{G^{99}}}\bigg)\!+\O\bigg(\big(\sqrt{G^{99}}\big)^{D\over 2}\, e^{- {2\pi c\over \sqrt{G^{99}}}}\bigg)~.
\end{aligned}
\label{MDs}
\ee}

If the cancellation of the NS-NS and RR characters in the Klein bottle amplitude~(\ref{KD}) makes the latter trivial, the torus contribution then needs prior consideration to be treated as $\A$ and $\M$. Modular invariance of the expression of $\T$ given in Eq.~(\ref{TD}) can be made explicit by writing the lattice of closed string zero modes in Lagrangian form, 
\be
\begin{aligned}
\T~=~ {1\over 2}\int_\F{d^2\tau\over \tau_2^{D+2\over 2}}&\,{1\over \eta^8\bar \eta^8} \,\half \sum_{a,b=0}^1(-1)^{a+b+ab}\, {\theta\big[{}^a_b\big]^4\over \eta^{4}}\, \half \sum_{\tilde a,\tilde b=0}^1(-1)^{\tilde a+\tilde b+\tilde a\tilde b}\, {\bar \theta\big[{}^{\tilde a}_{\tilde b}\big]^4\over \bar \eta^{4}} \\
&\; {\sqrt{\det G}\over \tau_2^{10-D\over 2}}\sum_{\vec l,\vec n}e^{-{\pi\over \tau_2}(l_I+n_I\bar \tau)G_{IJ}(l_J+n_J \tau)}(-1)^{l_9(a+\tilde a)+n_9(b+\tilde b)}~.
\label{TDL}
\end{aligned}
\ee
The last sign, which couples the spin structures $(a,b)$ and $(\tilde a,\tilde b)$ to the wrapping numbers $n_9, \tilde l_9$ of the worldsheet around the \SS direction $X^9$, is responsible for the spontaneous breaking of supersymmetry. One passes from Eq.~(\ref{TDL}) to Eq.~(\ref{TD}) by Poisson resummation over $l_I$, $I=D,\dots,9$. The above expression is explicitly modular invariant. Moreover, integration and  discrete sums over $l_9,n_9$ can be inverted in a suitable way, in order to ``unfold'' the fundamental domain $\F$. Schematically, we can write \cite{unfold}
\be
\begin{aligned}
\int_\F{d\tau_1d\tau_2}\sum_{l_9,n_9}f_{l_9,n_9}(\tau,\bar \tau)&~=~\int_\F{d\tau_1d\tau_2}f_{0,0}(\tau,\bar \tau)+\int_{-\half}^\half d\tau_1\int_0^{+\infty}{d\tau_2}\sum_{l_9\neq 0}f_{l_9,0}(\tau,\bar \tau) \\
&~=~\int_{-\half}^\half d\tau_1\int_0^{+\infty}{d\tau_2}\sum_{l_9}f_{l_9,0}(\tau,\bar \tau)~,
\end{aligned}
\ee
where in the second line we have used the fact that $f_{0,0}$ vanishes, due to supersymmetry. Turning back to the Hamiltonian form, Eq.~(\ref{TD}) can be written as
\be
\T~=~ {1\over 2}\int_{-\half}^\half d\tau_1\int_0^{+\infty}{d\tau_2\over \tau_2^{D+2\over 2}}\, {1\over \eta^8\bar \eta^8} \sum_{\vec m,\vec n'}\Big\{\big(V_8\bar V_8+S_8\bar S_8\big)\Lambda_{\vec m,(\vec n',0)}- \big( V_8\bar S_8+S_8\bar V_8\big)\Lambda_{\vec m+\vec a_S,(\vec n',0)} \Big\}~.
\label{TDunfo}
\ee
Integrating over $\tau_1$, which implements the level matching condition, and Poisson resumming over $m_9$, one obtains
\be
\begin{aligned}
\T~=~  (G^{99})^{D\over 2}\, {\Gamma\big({D+1\over 2}\big)\over \pi^{D+1\over 2}}\, 2\cdot8^2&\sum_{k,\tilde k\ge 0}c_kc_{\tilde k}\sum_{\vec m',\vec n'}\delta_{\vec m'\cdot\vec n'+k-\tilde k,0} \\
& \sum_{l_9}{ \cos\!\Big[2\pi(2l_9+1){G^{9i}\over G^{99}}m_i\Big]\over |2l_9+1|^{D+1}}\,  \Hc_{D+1\over 2}\Big(\pi|2l_9+1|{\M_{\T}\over \sqrt{G^{99}}}\Big)~, 
\end{aligned}
\ee
where we have defined
\be
\M_\T^2~=~P_i^L \hat G^{ij}P_j^L+k~=~P_i^R \hat G^{ij}P_j^R+\tilde k~.
\ee
It is not difficult to show that the KK towers such that $\M_\T$ is not $\O(1)$ satisfy $k=\tilde k=0$, $\vec m'=\vec n'=\vec 0'$, so that 
 \be
\begin{aligned}
\T~=~  \big(\sqrt{G^{99}}\big)^{D} \, {\Gamma\big({D+1\over 2}\big)\over \pi^{D+1\over 2}}\, 8\sum_{l_9}{16 \over |2l_9+1|^{D+1}} ~+~\O\bigg(\big(\sqrt{G^{99}}\big)^{D\over 2}\, e^{- {2\pi c\over \sqrt{G^{99}}}}\bigg)~.
\end{aligned}
\label{TDs}
\ee

In total, the effective potential, which combines all four worldsheet topologies, can be found in Eqs~(\ref{vtot}), (\ref{hatn}).


\paragraph{\em Effective potential at KK scales lower than \bm $M_s$~: }

To complete this section, we rederive the effective potential for arbitrary WL matrices~(\ref{WI}). This is done under an alternative assumption on the internal metric compared to the above analysis. Namely, we take all internal directions (in the original type~I picture) to be large, in string units,
\be
G_{II}\gg 1~, \qquad \mbox{(no sum on) }I=D,\dots,9~.
\ee 
This amounts to keeping all KK compactification scales lower than $M_s$ and all winding masses heavier than $M_s$. It is then convenient to apply a Poisson resummation on all internal momenta $m_I$, $I=D,\dots,9$, rather than on $m_9$ only. 

For the open string amplitudes~(\ref{ampA}) and~(\ref{ampM}), this is done by using Eq.~(\ref{eq:PTransform}).\footnote{However, we denote by $\vec l$ and not $\vec n$ the resummed indices to stress that we do not switch to the closed string channel. The $SO(8)$ affine characters remain $V_8$ and $S_8$ only, or their hatted counterparts.} Expanding the characters as before and utilising the definition of the function $\Hc_\nu$, we obtain 
\be
\begin{aligned}
\A&~=~ {8 \Gamma(5)\over \pi^5}\,  \sqrt{\det G}  \; \sum_{k\ge 0}c_k\sum_{\alpha,\beta}\sum_{\vec l}{e^{2i\pi \tilde l \cdot (\vec a_\alpha-\vec a_\beta)}\over (\tilde l_IG_{IJ}\tilde l_J)^5}\,\Hc_5\Big(\pi\sqrt{k\, \tilde l_I G_{IJ}\tilde l_J}\Big),\\
\M&~=~ -{8 \Gamma(5)\over \pi^5}\, \sqrt{\det G}  \; \sum_{k\ge 0}(-1)^kc_k\sum_{\alpha}\sum_{\vec l}{e^{4i\pi \tilde l \cdot \vec a_\alpha}\over (\tilde l_IG_{IJ}\tilde l_J)^5}\,\Hc_5\Big(\pi\sqrt{k\, \tilde l_I G_{IJ}\tilde l_J}\Big)~,
\end{aligned}
\ee
where $\tilde l$ is a vector whose last entry is odd, 
\be
\vec l~\equiv~(\vec l',l_9)\in \Z^{10-D}\quad \Longrightarrow\quad \tilde l\equiv (\vec l',2l_9+1)~.
\ee
By noting that the argument of $\Hc_5$ is $\O(\sqrt{G_{99}})$, unless it vanishes when $k=0$, we conclude that
\be
\begin{aligned}
\A+\M~=~  {8 \Gamma(5)\over \pi^5}\, \sqrt{\det G}  \; \sum_{\vec l}&{\big(\text{tr}\, (\W_D^{\tilde l_D}\cdots\W_9^{\tilde l_9})\big)^{2}-\text{tr}\, (\W_D^{2\tilde l_D}\cdots\W_9^{2\tilde l_9})\over (\tilde l_IG_{IJ}\tilde l_J)^5}\\
&\qquad\qquad\qquad\qquad~+~\O\big(\sqrt{\det G}\, G_{99}^{-{11\over 4}}e^{-2\pi\sqrt{G_{99}}}\big)~.
\end{aligned}
\ee
In the above result, the annulus contribution is formulated in terms of a squared trace by recalling that branes go in pairs with their mirrors, or are frozen at corners of the internal box. 

In the closed string sector, the torus amplitude provides the only contribution. At large internal directions, all winding modes in Eq.~(\ref{TD}) yield exponentially suppressed corrections. Hence, the Lagrangian form, Eq.~(\ref{TDL}), may be written 
\be
\begin{aligned}
\T~=~ {1\over 2}\int_\F{d^2\tau\over \tau_2^{D+2\over 2}}&\, {1\over \eta^8\bar \eta^8}\, \half \sum_{a,b=0}^1(-1)^{a+b+ab}\, {\theta\big[{}^a_b\big]^4\over \eta^{4}}\, \half \sum_{\tilde a,\tilde b=0}^1(-1)^{\tilde a+\tilde b+\tilde a\tilde b}\, {\bar \theta\big[{}^{\tilde a}_{\tilde b}\big]^4\over \bar \eta^{4}} \\
&\; {\sqrt{\det G}\over \tau_2^{10-D\over 2}}\sum_{\vec l}e^{-{\pi\over \tau_2}l_IG_{IJ}l_J}(-1)^{l_9(a+\tilde a)}~+~\O(e^{-\#\inf G_{II}})~,
\end{aligned}
\ee
where $\#=\O(1)$ is positive. Since even $l_9$ yields supersymmetric and therefore vanishing contributions, we can change $l_9\to 2l_9+1$. By noting that 
\be
\int_\F d\tau_1d\tau_2 \, e^{-{\pi\over \tau_2}(2l_9+1)^2G_{99}}(\, \cdots)~=~\int_{-\half}^\half d\tau_1\int_0^{+\infty}{d\tau_2} \, e^{-{\pi\over \tau_2}(2l_9+1)^2G_{99}}(\,\cdots)+\O(e^{-\#G_{99}})\,,
\ee
we obtain
\be
\begin{aligned}
\T&~=~ {\sqrt{\det G}\over 2}\int_{-\half}^\half d\tau_1\int_0^{+\infty}{d\tau_2\over \tau_2^{1+5}}\, {\theta_2^4\over \eta^{12}}\, {\bar \theta_2^4\over \bar \eta^{12}} \sum_{\vec l}e^{-{\pi\over \tau_2}\tilde l_IG_{IJ}\tilde l_J}~+~\O(e^{-\#\inf G_{II}})\\
&~=~{\Gamma(5)\over \pi^5}\,  \sqrt{\det G}\; 8\sum_{k\ge 0}c_k^2\sum_{\vec l}
{16\over (\tilde l_IG_{IJ}\tilde l_J)^5}\,\Hc_5\Big(2\pi\sqrt{k\, \tilde l_I G_{IJ}\tilde l_J}\Big)~+~\O(e^{-\#\inf G_{II}})\\
&~=~ {\Gamma(5)\over \pi^5}\, \sqrt{\det G}  \; 8\sum_{\vec l}{16\over (\tilde l_IG_{IJ}\tilde l_J)^5}~+~\O\big(\sqrt{\det G}\, G_{99}^{-{11\over 4}}e^{-4\pi\sqrt{G_{99}}}\big)~.
\end{aligned}
\ee

In total, the 1-loop effective potential~(\ref{Vdef}) then takes the final form \be
\begin{aligned}
\V~=~{\Gamma(5)\over \pi^{D+5}}\, {M_s^D\over 2^D} \sqrt{\det G} \;  4\sum_{\vec l}&{-16-\big(\text{tr}\, (\W_D^{\tilde l_D}\cdots\W_9^{\tilde l_9})\big)^{2}+\text{tr}\, (\W_D^{2\tilde l_D}\cdots\W_9^{2\tilde l_9})\over (\tilde l_IG_{IJ}\tilde l_J)^5}\\
&\qquad\qquad\qquad~+~\O\big(M_s^D\sqrt{\det G}\, G_{99}^{-{11\over 4}}e^{-2\pi\sqrt{G_{99}}}\big)~.
\end{aligned}
\label{VLtot}
\ee


\end{document}